\documentclass[useAMS,usenatbib]{mn2e}
\usepackage{graphicx}
\usepackage{amsfonts}
\usepackage{pifont}

\usepackage{color}
\definecolor{purple}{RGB}{160,32,240}

\def\lcdm{\mbox{$\Lambda$CDM}}
\def\gsmf{\mbox{GSMF}}
\def\shmr{\mbox{SHMR}}
\def\sfr{\mbox{SFR}}
\def\ssfr{\mbox{sSFR}}
\def\smar{\mbox{sMAR}}
\def\mar{\mbox{MAR}}

\def\msun{\mbox{M$_{\odot}$}}

\def\phih{\mbox{$\phi_{\rm h}$}}

\def\ms{\mbox{$M_*$}}
\def\mgas{\mbox{$M_{\rm gas, ISM}$}}
\def\fb{\mbox{$f_{\rm b}$}}
\def\xieff{\mbox{${\cal{E}}_{\rm eff}$}}
\def\dmsdt{\mbox{$\dot{M}_*$}}
\def\dmvirdt{\mbox{$\dot{M}_{\rm vir}$}}
\def\dmvirdyn{\mbox{$\dot{M}_{\rm vir,dyn}$}}
\def\mvir{\mbox{$M_{\rm vir}$}}
\def\Pcen{\mbox{$P(\ms|\mvir)$}}

\def\dmgravdt{\mbox{$\dot{M}_{\rm grav, ISM}$}}
\def\dmfalldt{\mbox{$\dot{M}_{\rm r, ISM}$}}
\def\dmoutdt{\mbox{$\dot{M}_{\rm out, ISM}$}}
\def\mdgasdt{\mbox{$\dot{M}_{\rm gas, ISM}$}}
\def\mzdt{\mbox{$\dot{M}_{\rm Z, ISM}$}}
\def\mz{\mbox{$M_{\rm Z, ISM}$}}
\def\etaout{\mbox{$\eta_{\rm w, ISM}$}}
\def\etafall{\mbox{$\eta_{\rm r, ISM}$}}

\def\zigm{\mbox{$Z_{\rm IGM}$}}
\def\zism{\mbox{$Z_{\rm ISM}$}}
\def\zout{\mbox{$Z_{\rm w, ISM}$}}
\def\zfall{\mbox{$Z_{\rm r, ISM}$}}

\def\dzismdt{\mbox{$\dot{Z}_{\rm ISM}$}}

\def\ltsima{$\; \buildrel < \over \sim \;$}    
\def\lesssim{\lower.5ex\hbox{\ltsima}}           
\def\gtsima{$\; \buildrel > \over \sim \;$}    
\def\grtsim{\lower.5ex\hbox{\gtsima}}           

\title[Stellar-Halo Accretion Rate Coevolution]
 {Is Main Sequence Galaxy Star Formation Controlled by Halo Mass Accretion?}
\author[]{Aldo Rodr\'iguez-Puebla$^1$\thanks{arodr104@ucsc.edu}, Joel R. Primack$^2$, Peter Behroozi$^3$, and S. M. Faber$^4$ \\
$^1$Department of Astronomy and Astrophysics, University of California, Santa Cruz, CA 95064, USA \\
$^2$Department of Physics, University of California, Santa Cruz, CA, 95064, USA \\
$^3$Space Telescope Science Institute, Baltimore, MD 21218, USA \\
$^4$UCO/Lick Observatory, Department of Astronomy and Astrophysics, University of California, Santa Cruz, CA 95064, USA \\
}

\begin{document}

\label{firstpage}

\maketitle

\begin{abstract}
The galaxy stellar-to-halo mass relation (SHMR) is nearly time-independent for $z < 4$. We therefore construct a time-independent SHMR model for central galaxies, wherein the in-situ star formation rate (SFR) is determined by the halo mass accretion rate (MAR), which we call Stellar-Halo Accretion Rate Coevolution (SHARC). We show that the $\sim0.3$ dex dispersion of the halo MAR matches the observed dispersion of the SFR on the star-formation main sequence (MS). In the context of ``bathtub"-type models of galaxy formation, SHARC leads to mass-dependent constraints on the relation between SFR and MAR. Despite its simplicity and the simplified treatment of mass growth from mergers, the SHARC model is likely to be a good approximation for central galaxies with $M_*=10^9- 10^{10.5}M_\odot$ that are on the MS, representing most of the star formation in the Universe. SHARC predictions agree with observed SFRs for galaxies on the MS at low redshifts, agree fairly well at $z\sim4$, but exceed observations at $z\grtsim4$. Assuming that the interstellar gas mass is constant for each galaxy (the ``equilibrium condition'' in bathtub models), the SHARC model allows calculation of net mass loading factors for inflowing and outflowing gas. With assumptions about preventive feedback based on simulations, SHARC allows calculation of galaxy metallicity evolution. If galaxy SFRs indeed track halo MARs, especially at low redshifts, that may help explain the success of models linking galaxy properties to halos (including age-matching) and the similarities between two-halo galaxy conformity and halo mass accretion conformity.
\end{abstract}

\begin{keywords}
cosmology: theory -- galaxies: halos -- galaxies: evolution -- methods: N-body simulations -- methods: luminosity function, mass function
\end{keywords}

\section{Introduction}

In the cold dark matter paradigm, galaxies form in dark matter halos.  As cosmological 
simulations such as Millennium \citep{Millennium,MillenniumII} and Bolshoi \citep{Klypin+2011} resolved dark matter halos increasingly well, it has been a major goal to use such
simulations to improve our understanding of the connection between halos and the galaxies that they host.  Abundance matching \citep{Kravtsov+2004,ValeOstriker04} -- which in its most basic form is just rank ordering galaxies by their stellar mass and assigning them to halos ranked by mass or peak circular velocity -- leads to predictions of the galaxy autocorrelation functions for both bright and faint galaxies that are in excellent agreement with observations \cite[e.g.,][and references therein]{Conroy+2006,Klypin+2011,RDA12,Reddick+2013}.  Abundance matching taking into account galaxy star formation rates also allows calculation of the typical relationship between the mass of dark matter halos and the stellar mass of the hosted galaxies \citep[e.g.][]{Moster+2013,Behroozi+2013a}.  The resulting stellar-to-halo mass relation (SHMR) is remarkably similar at all redshifts between 0 and 4.  
This is consistent with the assumption that the average virial star formation efficiency (the star formation rate divided by the halo baryon accretion rate) is only a function of halo mass and not redshift from $z=4$ to the present epoch \citep{Behroozi+2013c}.  
This motivates us to develop a simple model in which the mass accretion rate (MAR) of dark matter halos determines the star formation rate (SFR) of their host galaxies.  
We call this the Stellar-Halo Accretion Rate coevolution (SHARC) assumption.  Note that it is also possible to develop
different galaxy-halo coevolution models that could also satisfy the time-independent SHMR by 
correlating SFRs to other halo assembly properties 
(e.g., halo formation time) but SHARC is a particularly simple one.

The mass accretion rate of dark matter halos depends on the precise definition of the halos, which has been called into question in several recent papers.
\citet{Diemer+2013} argued that much of the mass evolution of dark matter halos is an artifact caused by the changing radius of the halo, a phenomenon that they call ``pseudo-evolution."  In this paper we define the radius of the halo as the radius $R_{\rm vir}$ that encloses an average density 
$\Delta_{\rm vir} \rho_{\rm m}$, where $\rho_{\rm m}$ is the mean matter density of the universe 
$\Omega_{\rm M} \rho_{\rm c}$, $\rho_{\rm c}$ is critical density, and the redshift-dependent virial overdensity $\Delta_{\rm vir}(z)$ is given by the spherical collapse model \citep{BryanNorman}.  Other popular definitions are $R_{\rm 200m}$ and $R_{\rm 200c}$, corresponding to enclosed densites of $200 \rho_{\rm m}$ and $200 \rho_{\rm c}$ respectively.  For all these definitions, the rapid drop in background density as $z$ decreases is the main cause of the increase in halo virial radius and therefore a main cause of the increase in the enclosed mass, while the dark matter distribution in the interior of the halo hardly changes at low redshift \citep{Prada+2006,Diemand+2007,Cuesta+2008}.  Recently \cite{More+2015} proposed that the best physically-based definition of halo radius is the ``splashback radius'' $R_{\rm sp} \approx 2 R_{\rm 200m}$, where there is typically a sharp drop in the density; using this definition, there is actually more halo mass increase than for $R_{\rm vir}$, $R_{\rm 200m}$, or $R_{\rm 200c}$.  

What is actually relevant to star formation of the central galaxy in the halo is the amount of gas that enters the halo and eventually reaches its central regions.  \citet{WetzelNagai2015} have used adaptive refinement tree (ART) hydrodynamic galaxy simulations with a best resolution of about 0.5 kpc to show that infalling gas decouples from dark matter starting at about $2 R_{\rm 200m}$ and roughly tracks the 
growth of $M_{\rm 200m}$.
Thus, they argue that pseudo-evolution is relevant to the accretion of dark matter, but not to that of gas.  
Further evidence that the gas falling into the central regions of  halos roughly tracks the halo mass accretion rate is provided by \citet{Dekel+2013}, who used a suite of ART hydrodynamic zoom-in galaxy simulations with $M_{\rm vir}/10^{12} M_\odot = 0.1$ to 2 at $z=2$ and best resolution of 35 pc, and found that the gas inflow rate is proportional to the halo mass accretion rate, with about half the gas penetrating to $0.1 R_{\rm vir}$ at redshifts $z=4$ to 2, and with the fraction increasing from $z=2$ to 1.  Analysis of a subsequent suite of ART hydrodynamic zoom-in galaxy simulations with better resolution and feedback showed that these simulated star-forming galaxies grow stellar mass at the same rate as the specific halo mass increase \citep[][Tacchella et al., in prep.]
{Zolotov+2015}. A similar result has been reported in \citet{GS+2014} for galaxies formed in halos of 
$\mvir=2-3\times10^{10}\msun$. 

\begin{figure*}
	\vspace*{-150pt}
	\includegraphics[height=6in,width=6in]{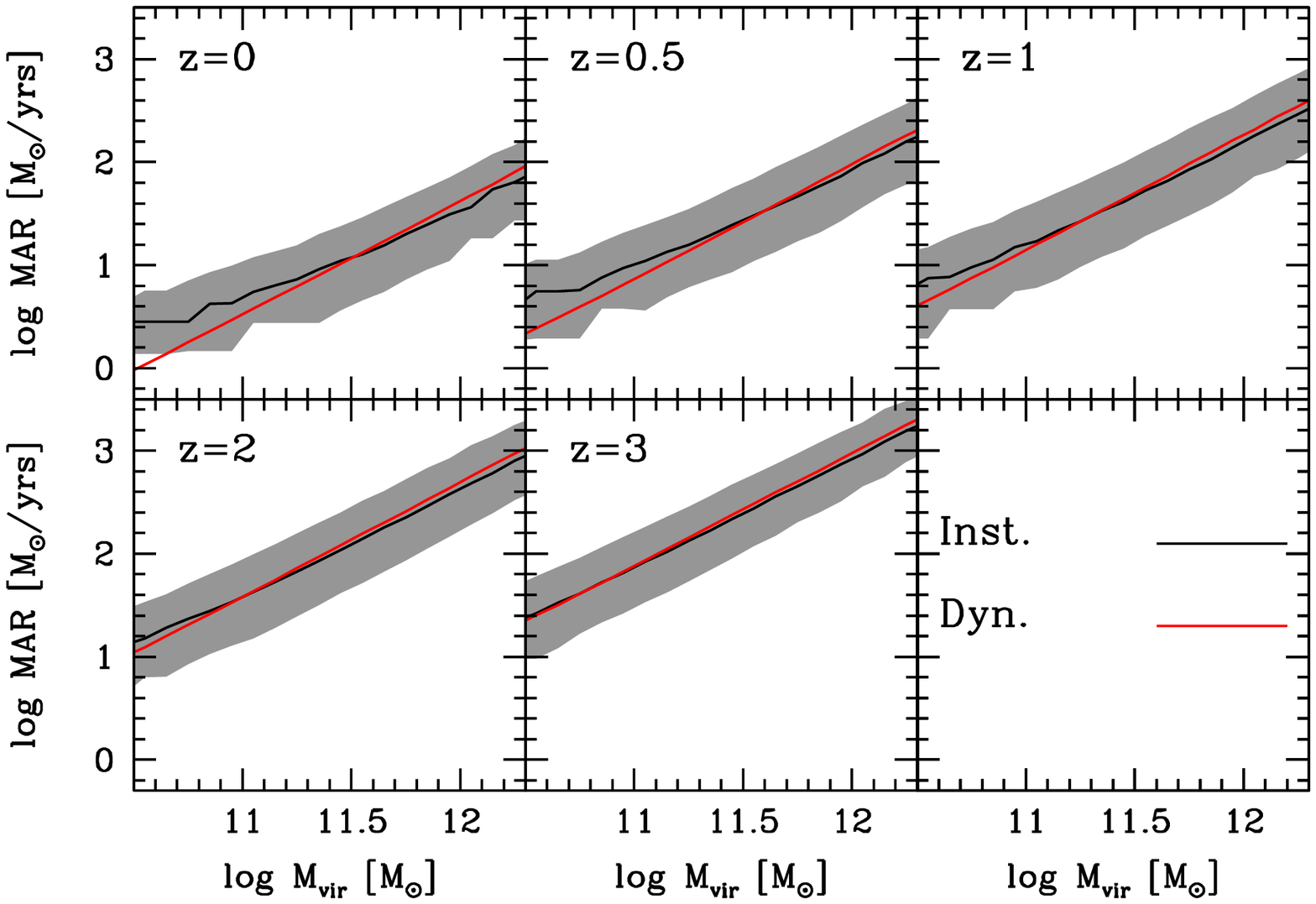}
		\caption{Halo mass accretion rates (MARs) from $z=0$ to $z=3$, from the Bolshoi-Planck simulation.  The instantaneous
		rate is shown in black, and the dynamically time averaged rate in red.  The gray band
		is the $1\sigma$ (68\%) range of the instantaneous mass accretion rates. All the slopes are approximately
		the same $\sim1.1$ both for \dmvirdt\ and \dmvirdyn. 
 	}
	\label{HMAR}
\end{figure*}

Star-forming galaxies are known to show a tight dependence of SFR on stellar mass, which is known as the ``main sequence" of galaxy formation \citep{Salim+2007,Noeske+2007,Elbaz+2007,Daddi+2007}, in analogy with the tight dependence on stellar mass of the properties of stars on the stellar main sequence.  Dark matter halos also have a mass accretion rate that is roughly proportional to their mass, and we show in \S\ref{Simulation} of this paper that the dispersion of the halo mass accretion rate at a given halo mass is 0.3 to 0.4 dex, similar to the dispersion of the SFR on the main sequence. It was this equality that originally motivates us to examine more closely the connection between mass accretion and star formation.

Our analysis is limited to the connection between distinct halos and central galaxies only. 
One reason is that subhalos lose mass via tidal stripping, resulting 
in negative values of accretion rates \citep[see, e.g.,][]{vandenBosch+2005}. Satellite galaxies are also affected in other ways by their proximity to central galaxies.  Therefore, studying the connection 
between subhalo mass accretion and satellite SFRs is beyond the scope of this paper. In order to focus on main-sequence galaxies, we also discuss mainly dark matter halos of masses $10^{11}$ to $10^{12} M_\odot$.

In \S\ref{model} we will derive the SHMR for all SDSS central galaxies based on the \citet{Yang+2012}
and \citet{RP+2015}. Presently available data allows this to be done only for $z\sim0$. In this paper,
we make the simple assumption that this SHMR is valid at all redshifts. The SHARC assumption will allow us to deduce
the SFR for every halo in the Bolshoi-Planck simulation. When we compare these predictions with observations
we find that they are in pretty good agreement from $z=0$ to $\sim4$, both for the SFR and its dispersion. 

Star formation is regulated by a complex interaction between gas inflows and outflows.  
Models that describe the basic processes that govern gas inflows and outflows and star formation in galaxies
 are called ``bathtub" models in the literature
\citep{Bouche+2010,Dave+2011,Dave+2012,Krumholz+2012,Dekel+2013,Lilly+2013,
DekelMandelker2014,Forbes+2014,Feldmann2015,Mitra+2015}. 
Such models consider both ``equilibrium'' conditions when the amount of gas in the interstellar medium (the ``bathtub'') is constant because star formation equals net inflow, and non-equilibrium situations when the bathtub is filling or emptying.  
In the even simpler model in this paper we assume equilibrium at all times.
We call this the Equilibrium and SHARC model, or E+SHARC.
As is shown in \S\ref{EQC}, the  time-independent \shmr\ is compatible with the equilibrium condition.

%
 
This paper is organized as follows: \S\ref{model}  describes the dark matter simulation used here and how we connect central galaxies to host halos.  There we also make the simplifying assumption that the stellar-to-halo mass relation for central galaxies
on the main sequences is independent of redshift
and show how this allows us to infer SFRs from halo mass accretion rates, i.e., the SHARC assumption.  In \S\ref{EQC} we explore a very simple bathtub model, which we assume for simplicity to be in equilibrium at all times -- i.e., the gas mass is constant -- in order to identify and understand the conditions that are satisfied by the equilibrium time-independent \shmr\ model, i.e., the  E+SHARC assumption.
We show how net gas infall is connected to preventive feedback.  Assuming a power-law preventive feedback for $M_{\rm vir} \grtsim 10^{12} M_\odot$ (representing virial shock heating of in-falling gas and the effects of AGN), we deduce mass-loading factors and their dispersion as a function of halo mass and redshift.  In \S\ref{SFR_section} we deduce the SFRs and their dispersion implied by our model, and compare with observations of the SFRs and dispersion on the star-forming main sequence. Not surprisingly, we find that our simple model  does not correctly predict the SFR at high redshifts $z\grtsim 4$, showing
that the SHMR should change above $z\sim4$. 
However, from $z=4$ to 0, the SFR predictions from the time-independent \shmr\ model are in better agreement with the observed SFRs on the main sequence, especially if we use the most recent observations.  In \S\ref{CSFR_SMF} we compare the cosmic star formation rate density and galaxy stellar mass function with observations up to $z\sim6$, again finding that the SHMR
assumption is disfavored at high redshift.  In \S\ref{Metals_EQC} we calculate the metallicity of the interstellar medium given by our
E+SHARC model and compare with observations, yet again finding that this model fails at high redshift.  Finally, 
\S\ref{Conclusions} summarizes our conclusions, discusses their implications, and describes ways to increase the generality 
of the simplified model considered here. 

We adopt cosmological parameters
$\Omega_\Lambda=0.693$, $\Omega_{\rm M}=0.307$, $\Omega_{\rm bar}=0.048$, $h=0.678$, $n_s=0.96$, 
and $\sigma_8=0.829$, consistent with recent results from the Planck Collaboration \citep{Planck13,Planck15}.  These are the parameters used in the Bolshoi-Planck simulation \citep[][Rodriguez-Puebla et al. 2015 in prep.]{Klypin+2014}, on which our results here are based;
as noted above, our halo masses are defined using the spherical overdensity criterion of \cite{BryanNorman}.  We also assume a \cite{Chabrier2003} IMF. Finally, Table \ref{acronyms} lists all the acronyms
used in this paper.

\begin{table}
	\caption{List of acronyms used in this paper.}
		\begin{center}
			\begin{tabular}{| l l}
			\hline
			\hline
			ART & Adaptive Refinement Tree (simulation code)\\
			CSFR & Cosmic Star Formation Rate\\
			IMF	& Initial Mass Function\\
			ISM & Interstellar Medium\\
			GSMF & Galaxy Stellar Mass Function\\
			MAR & Mass Accretion Rate, \dmvirdt \\
			SHARC & Stellar-Halo Accretion Rate Coevolution \\
			E+SHARC & Equilibrium+SHARC \\
			SDSS & Sloan Digital Sky Survey \\
			SFR & Star Formation Rate \\
			SHMR & Stellar-to-Halo Mass Relation \\
			sMAR & Specific Mass Accretion Rate,  \dmvirdt/\mvir \\
			sSFR & specific Star Formation Rate, \sfr/\ms \\
			\hline
			\end{tabular}
		\end{center}
	\label{acronyms}
\end{table}%

\section{Stellar-Halo Accretion Rate Coevolution (SHARC)}
\label{model}

\subsection{The Simulation}
\label{Simulation}

We generate our mock galaxy catalogs based on the N-body \textit{Bolshoi-Planck} simulation \citep{Klypin+2014}.
The Bolshoi-Planck simulation is based on the 
\lcdm\ cosmology with parameters consistent with the latest results from the
Planck Collaboration \citep{Planck15} and run using the Adaptive Refinement
Tree code \citep[ART][]{Kravtsov+1997,Gottloeber+2008}. The Bolshoi-Planck simulation
has a volume of $(250 h^{-1} {\rm Mpc})^3$ and contains $2048^3$ particles of
mass $1.9\times 10^8$ \msun. Halos/subhalos and their merger trees
were calculated with the phase-space temporal halo finder ROCKSTAR
\citep{Behroozi+2013b,Behroozi+2013d}. Halo masses were defined using 
spherical overdensities according to the redshift-dependent virial overdensity $\Delta_{\rm vir}(z)$ 
given by the spherical collapse model \citep{BryanNorman}, with $\Delta_{\rm vir} = 178$ 
for large $z$ and $\Delta_{\rm vir} = 333$ at $z=0$ with our $\Omega_{\rm M}$.  
Like the Bolshoi simulation \citep{Klypin+2011}, Bolshoi-Planck is complete down to 
halos of maximum circular velocity $v_{\rm max}\sim55$ km/s.  

In this paper, we calculate instantaneous halo mass accretion rates from the Bolshoi-Planck 
simulation, as well as halo mass accretion rates averaged over the dynamical time (\dmvirdyn), defined as
	\begin{equation}
		\big\langle\frac{d\mvir}{dt}\big\rangle_{\rm dyn} = \frac{\mvir(t) - \mvir(t-t_{\rm dyn})}{t_{\rm dyn}}.
	\end{equation}
The dynamical time of the halo is $t_{\rm dyn}(z) = [G \Delta_{\rm vir}(z) \rho_{\rm m}]^{-1/2}$,
which is $\sim 20\%$ of the Hubble time.  
Simulations \citep[e.g.,][]{Dekel+2009} suggest that most star formation results from cold gas flowing inward at about the virial velocity -- i.e., roughly a dynamical time after the gas enters.  As instantaneous accretion rates for distinct halos near clusters can also be negative \citep{Behroozi14}, using time-averaged accretion rates allows galaxies in these halos to continue forming stars.

\begin{figure}
	\vspace*{-80pt}
	\hspace*{-20pt}
	\includegraphics[height=3.8in,width=3.8in]{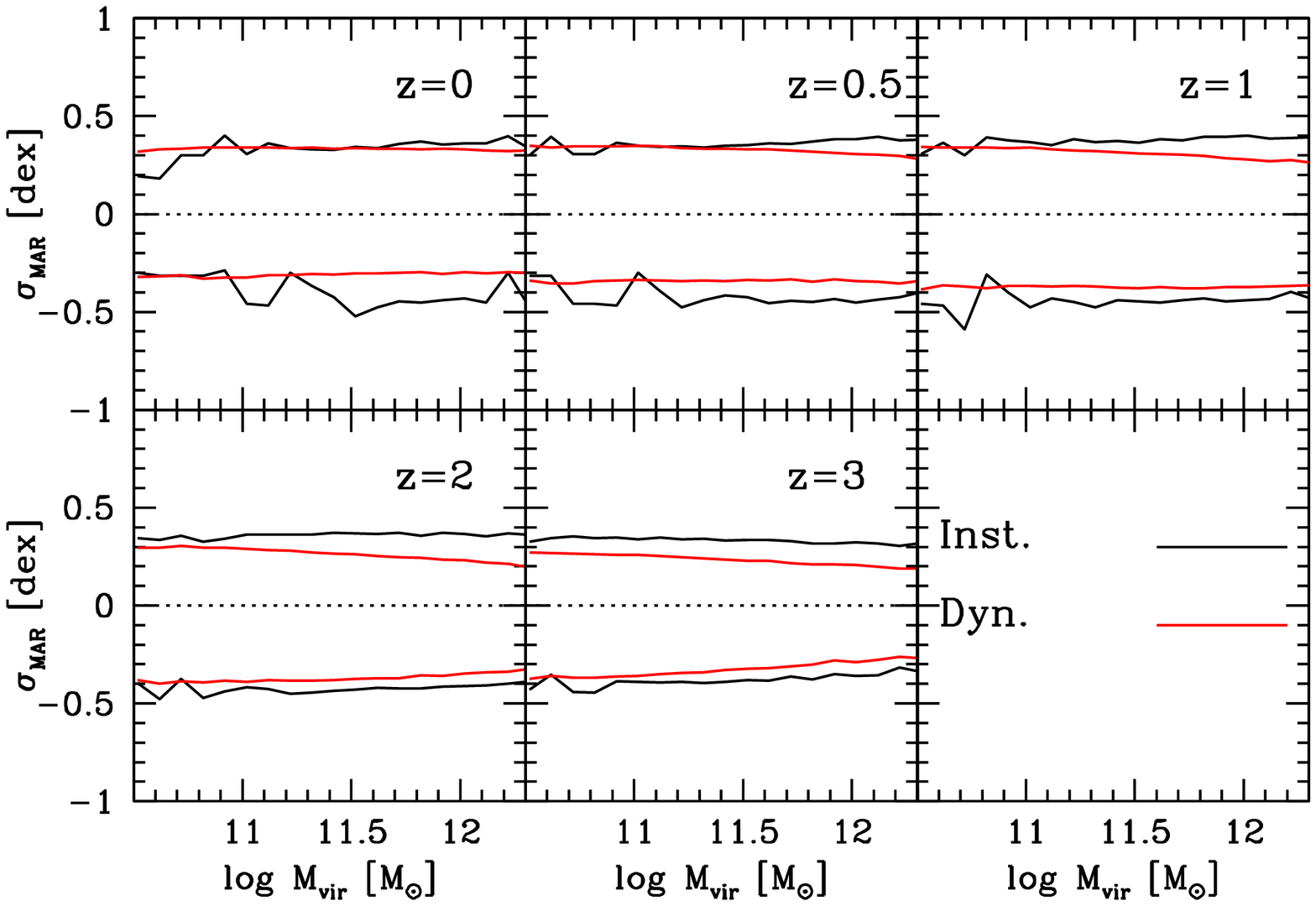}
		\caption{Scatter of halo mass accretion rates from $z=0$ to $z=3$ from the Bolshoi-Planck simulation.  As in Fig.\ \ref{HMAR}, scatter for the instantaneous
		rate is shown in black, and that for the dynamically time averaged rate in red.
 	}
	\label{HMARscatter}
\end{figure}

Figure~\ref{HMAR} shows the instantaneous and the dynamical-time-averaged halo mass accretion rates as a function of halo mass and redshift, 
and Figure~\ref{HMARscatter} shows their respective scatters. Even before converting halo accretion rates into star formation rates (\S \ref{SHARC}), it is evident that both the slope and dispersion in halo mass accretion rates are already very similar to that of galaxy star formation rates on the main sequence.


\subsection{Connecting Galaxies to Halos}
\label{G-H}

\begin{figure}
	\vspace*{-95pt}
	\hspace*{-20pt}
	\includegraphics[height=6.5in,width=6.5in]{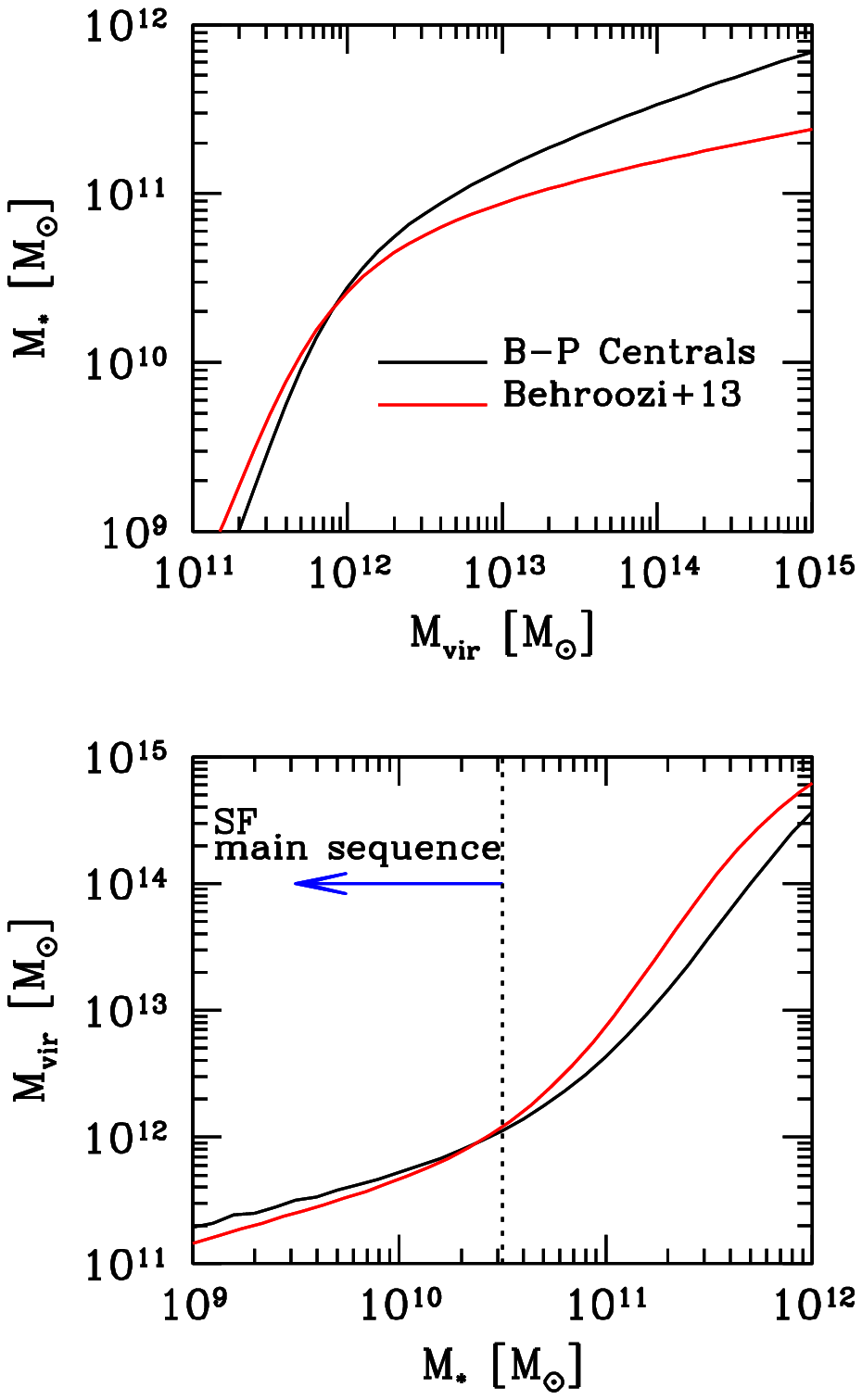}
		\caption{{\bf Upper Panel:} Stellar-to-halo mass relation (SHMR) for SDSS galaxies. 
		The red curve is for all SDSS galaxies, from \citet{Behroozi+2013a} abundance matching using the Bolshoi simulation.
		The black curve is for SDSS central galaxies, using the abundance matching method of \citet{RAD13} applied to the Bolshoi-Planck simulation.  The latter is what we use in the present paper, where we restrict attention to central galaxies.
		{\bf Bottom Panel:} 
		Halo-to-stellar mass relations. The dotted vertical line and the 
		blue arrow indicate that galaxies below $\ms=10^{10.5}\msun$
		are considered as main sequence galaxies, while some higher-mass galaxies are not on the main sequence. 
		}
 		 \label{f3}
\end{figure}

The abundance matching technique is a simple and powerful
statistical approach to connecting galaxies to halos. In its most
simple form, the cumulative halo and subhalo mass function\footnote{Typically defined at the time of subhalo accretion.}
and the cumulative galaxy stellar mass function (\gsmf)
are matched in order to determine the mass relation between halos and
galaxies. In order to assign galaxies to halos in the Bolshoi-Planck
simulation, in this paper we use a more general procedure for 
abundance matching. Recent studies have shown that the mean
stellar-to-halo mass relations (\shmr) of central and satellite galaxies are 
slightly different, especially at lower masses where satellites
tend to have more stellar mass compared to centrals of the same halo mass 
\citep[for a more general discussion see][]{RDA12,RAD13,Reddick+2013,Watson+2013,Wetzel+2013}.
Since we are interested in studying the connection between halo mass
accretion and star formation in central galaxies, for our analysis we derive the 
\shmr\ for central galaxies only. 
  
We model the \gsmf\ of central galaxies by defining $\Pcen$ as 
the probability distribution function that a distinct halo of mass \mvir\ hosts
a central galaxy of stellar mass  $\ms$.  
Then the \gsmf\ for central galaxies as a function of stellar mass
is given by
\begin{equation}
\phi_{*, \rm cen}(\ms)=\int^{\infty}_{0}\Pcen\phih(\mvir)d\mvir.
\end{equation}
Here, $\phih(\mvir)$ is the halo mass function and \Pcen\ is a log-normal distribution assumed to
have a scatter of $\sigma_c=0.15$~dex independent of halo mass.
Such a value is supported by the analysis of large group catalogs
\citep{Yang+2009b,Reddick+2013}, studies of the kinematics of satellite galaxies  \citep{More+2011},
as well as clustering analysis of large samples of galaxies \citep{Shankar+2014,RP+2015}.
Note that this scatter, $\sigma_c$, consists of an intrinsic component and a 
measurement error component. At $z=0$, most of the scatter appears to be intrinsic, but
that becomes less and less true at higher redshifts \citep[see, e.g.,][]{Behroozi+2010,Behroozi+2013a,Leauthaud+2012,Tinker+2013}.
Here, we do not deconvolve to remove measurement error, as most of the observations
that we will compare to include these errors in their measurements. 
 
As regards the \gsmf\ of central galaxies, we here use the results reported in \citet{RP+2015}. 
In a recent analysis of the SDSS DR7, \citet{RP+2015} derived
the total, central, and satellite \gsmf\ 
for stellar masses from $\ms=10^{9}\msun$ to $\ms=10^{12}\msun$
based on the NYU-VAGC \citep{Blanton+2005} and using the $1/V_{\rm max}$ estimator. The membership (central/satellite) 
for each galaxy was obtained from an updated version of the \citet{Yang+2007} group catalog 
presented in \citet{Yang+2012}.  The corresponding SHMR is shown as the black curve
in Figure~\ref{f3}, and the SHMR for all galaxies from \citet{Behroozi+2013c} is shown as the red curve.
The difference between the two curves for halo masses lower than $\mvir\sim10^{12}\msun$ reflects
the fact that the \shmr\ of centrals and satellite galaxies are 
slightly different as mentioned above. At halo masses higher than $\mvir\sim10^{12}\msun$ ,
this difference is primarily due to the differences between the $\gsmf$s used to derive these
SHMRs, \citealp{Behroozi+2013b} used \citet{Moustakas+2013}. 
When comparing both $\gsmf$s we find that the high mass-end 
from \citet{RP+2015} is significantly different to the one derive in \citet{Moustakas+2013}. 
In contrast, when comparing \citet{RP+2015}  \gsmf\ with \citet{Bernardi+2010}  we
find an excellent agreement, for a more general discussion see \citet{RP+2015}. 
In less degree, we also find that the different values employed for the scatter of the
SHMR explain these differences.  
 
\subsection{Inferring Star Formation Rates From Halo Mass Accretion Rates}
\label{SHARC}

\begin{figure*}
	\vspace*{-360pt}
	\hspace*{-10pt}
	\includegraphics[height=8in,width=8in]{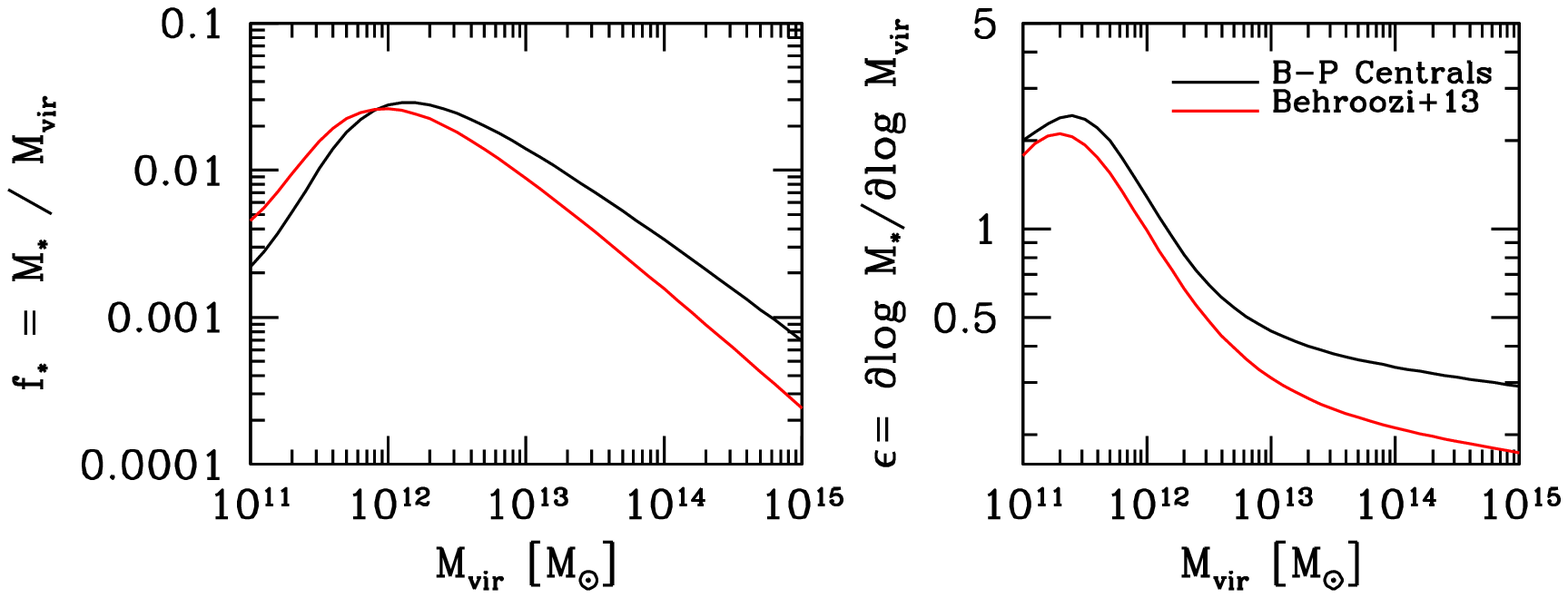}
		\caption{{\bf Left Panel:} Stellar-to-halo mass ratio for SDSS galaxies. 
		{\bf Right Panel:} Star formation efficiency.  
		As in Figure~\ref{f3}, the red curves are for all SDSS galaxies, from \citet{Behroozi+2013a},
		and the black curves are for SDSS central galaxies only for Bolshoi-Planck simulation.
		 	}
			\label{f4}
\end{figure*}

A number of recent studies exploring the \shmr\ at different redshifts 
have found that it evolves only slowly with time \citep[see, e.g.,][and references
therein]{Leauthaud+2012,Hudson+2014,Behroozi+2013a}. For example,
based on the observed evolution of the \gsmf, the
star formation rate \sfr, and the cosmic star formation rate, 
\citet{Behroozi+2013a}  showed that this is the case at least up to $z = 4$ (cf.\
possible increased evolution at $z>4$; \citealt{Behroozi15,Finkelstein15}). 
Moreover, \citet{Behroozi+2013c} showed that assuming a time-independent 
ratio of galaxy specific star formation rate (\ssfr) to host halo specific mass accretion rate (\smar), 
defined as the star formation efficiency $\epsilon$, simply explains the cosmic star formation rate since $z=4$.   
If we assume a time-independent \shmr, the star formation efficiency is
the slope of the \shmr,
	\begin{equation}
		\epsilon=\frac{\dmsdt/\ms}{\dmvirdt/\mvir}=\frac{\partial\log\ms}{\partial\log\mvir}.
		\label{SF_eff}
	\end{equation}
This equation simply relates galaxy 
$\sfr$s to their host halo $\mar$s without requiring knowledge of the underlying physics.
(This is the main difference between the equilibrium solution we present below and previous ``bathtub" models.)
Our primary motivation here is to understand whether halo $\mar$s are responsible for the
mass and redshift dependence of the $\sfr$ main sequence and its scatter.  Similar models have been
explored in the past for different purposes, including generating mock catalogs \citep{Popp15} and
understanding the different clustering of quenched and star-forming galaxies \citep{Becker2015}.

Using halo $\mar$s, we operationally infer galaxy $\sfr$s as follows. 
Let $\ms = \ms (\mvir(t),t) $
be the stellar mass of a central galaxy formed in a halo of mass $\mvir(t)$ at time $t$.   
In a time-independent \shmr, the above reduces to $\ms=\ms(\mvir(t))$. 
From this relation the change of stellar mass in time is simply 
	\begin{equation}
		\frac{d\ms}{dt} = f_* \frac{\partial\log\ms}{\partial\log\mvir} \frac{d\mvir}{dt},
			\label{dmsdt}
	\end{equation}
where $f_*=\ms/\mvir$ is the stellar-to-halo mass ratio. Equation (\ref{dmsdt}) implies  
stellar-halo accretion rate coevolution, SHARC.
The left panel of Figure \ref{f4} shows the resulting stellar-to-halo mass ratio, $f_*$, derived for SDSS central galaxies (see Section \ref{G-H}). Consistent with previous studies, we find that $f_*$ has a maximum of $\sim0.03$ at $\mvir\sim10^{12}\msun$, and it decreases at both higher and lower halo masses. The product $f_* \times \epsilon = dM_*/dM_{\rm vir}$ will be shown as
the black curves in Figure \ref{feedback_param} below.

In the more general case $\ms=\ms(\mvir(t),z)$, equation (\ref{dmsdt}) generalizes to
	\begin{equation}
		\frac{d\ms}{dt} =  \frac{\partial\ms(\mvir(t),z)}{\partial\mvir} \frac{d\mvir}{dt}+\frac{\partial\ms(\mvir(t),z)}{\partial z} \frac{dz}{dt},
			\label{full_dmsdt}
	\end{equation}
where the first term is the contribution to the SFR from halo MAR and the second term is the change in the
SHMR with redshift. 
Although in this paper we assume a constant SHMR, the formalism that we describe below applies to this more
general case. 

The relation between stellar mass growth and observed star formation rate is given by
	\begin{equation}
		\sfr = \dmsdt / (1 - R ),
	\end{equation}
where $R$ is the fraction of mass that is returned as gaseous material 
into the interstellar medium, ISM, from stellar winds and short lived stars.
In other words, $1-R$ is
the fraction of the change in stellar mass that is kept as long-lived. 
Here we make the instantaneous recycling 
approximation, with $R=R(t)$ as derived in \citet[][\S2.3]{Behroozi+2013a} and consistent with the \citet{Chabrier2003} IMF.  (In our model, for simplicity we take $t$ to be the cosmic time since the Big Bang.)

\subsection{Star Formation Efficiency}

Note that the star formation efficiency $\epsilon$, equation (\ref{SF_eff}), 
also quantifies galaxy stellar versus halo
mass growth. The right panel in Figure \ref{f4} shows the star formation efficiency 
as a function of halo mass. Several features in this figure are worth discussing. 
As has been long established, the star formation efficiency decreases significantly from 
$\mvir\sim10^{11}\msun$ to $\mvir\sim10^{14}\msun$, implying strong differences between galaxy and halo mass growth. 
Low-mass halos gain mass more slowly than low-mass galaxies. For high-mass halos, this trend is 
inverted: high-mass halos grow faster than high-mass galaxies. This is commonly called ``downsizing''
\citep[see][and references therein]{Fontanot+2009,Conroy+2009,Firmani+2010a}. 
Secondly, Milky-Way sized halos, $\mvir\sim10^{12}\msun$, have a star formation
efficiency of $\epsilon\sim1$.  Note that, for $\epsilon=1$, galaxy mass growth becomes linearly proportional
to the host halo's mass growth ($\ms\propto\mvir$). 

It is useful to rewrite the $\sfr$ as a function of the 
star formation efficiency, 
	\begin{equation}
		\sfr= f_*\times\epsilon\times \dmvirdt / (1 - R).
		\label{SFR_model}
	\end{equation}
This robust new result of the SHARC assumption can also be written as
	\begin{equation}
	F(M_{\rm vir}) \equiv \frac{\sfr}{f_b \dmvirdt } = \frac{f_* \times  \epsilon}{f_b (1-R)},
	\label{virial_SF}
	\end{equation}
which can be termed the ``virial star-formation efficiency."  Here 
the universal baryon fraction is defined as $\fb= \Omega_{\rm bar} / \Omega_{\rm M}$, and
 has a value of $\fb=0.156$ with the Planck cosmological parameters adopted for this paper.

While in the analysis described above $\sfr$s are based on instantaneous halo
mass accretion rates \dmvirdt, we also derive relations when using mass accretion rates averaged over
a dynamical time scale, \dmvirdyn. Specifically, in equation (\ref{SFR_model}) we 
substitute \dmvirdt\ by \dmvirdyn, given by equation (1).

We only expect equations (\ref{SFR_model}) and (\ref{virial_SF}) to apply to star-forming galaxies
on the main sequence. For our SDSS calibration sample this includes galaxies with $\ms=10^{9}-10^{10.5}\msun$, so it is this mass range, shown by the dotted line a blue arrow in Figure \ref{f3}, where we use the SHARC assumption in the rest of this paper. 
Note that above these masses quenched galaxies are detached from the mass accretion-star formation correlation. 
	
\subsection{Impact of Mergers}	
	\label{Mergers}
Both in-situ star formation and galaxy mergers can contribute to the stellar mass growth of 
galaxies. But most mergers of galaxies with $M_{\rm vir} > 10^{11} M_\odot$ at low redshift 
are dry mergers, so in inferring the stellar mass growth we should not include halo mass 
growth due to dry mergers. In addition to stars, merging halos may also contain diffuse circumgalactic medium. This will only contribute to the growth of the total baryonic content of the halo but not to the stellar mass growth of the central galaxy. 
  We do this  by multiplying the total stellar mass growth, \dmsdt, inferred naively from the halo mass growth, by the fraction of stellar mass growth of central galaxies that comes from star formation, $f_{\rm SFR}$. Specifically, we calculate in-situ
star formation in central galaxies only as 
\begin{equation}
\sfr =\frac{\dmsdt}{1 - R } f_{\rm SFR} (\mvir).
\end{equation}
We parameterize $f_{\rm SFR} (\mvir)$ as a function of halo mass and redshift following \citet[][equations 19-21 and Figure 10]{Behroozi+2013a}.

\section{SHARC + Bathtub: Equilibrium Assumption, Inflows, and Outflows}
\label{EQC}
  
The most important result from the previous section is that galaxy $\sfr$s
can be derived directly from halo $\mar$s 
if the \shmr\ is time-independent, i.e., the SHARC model. In this section, we explore 
a very simple gas-regulated model in order to identify and understand 
the conditions that are satisfied by the SHARC model. As we will show later,
the time-independent \shmr\ is compatible with the equilibrium condition
where galaxies are  regulated only between inflow, outflow and star-formation.

\subsection{Gas Equation}

We begin by defining the gas equation that describes the change of cold gas mass
in the interstellar medium (ISM) of a galaxy. 
The model described in the following section is a simple version of previous models 
discussed in the literature, sometimes called ``bathtub" models
\citep{Bouche+2010,Dave+2011,Dave+2012,Krumholz+2012,Dekel+2013,Lilly+2013,
DekelMandelker2014,Forbes+2014,Feldmann2015,Mitra+2015} 
describing the basic processes that govern the different components in galaxies. 

The change of the total cold gas mass within the ISM of a galaxy, \mdgasdt, is the result 
of the following physical mechanisms:
\begin{description}
\item[(i)] The rate at which the cosmological baryonic inflow material will reach the ISM of the 
galaxy, \dmgravdt. This process is assumed to be related to the gravitational
structure formation of the halo, i.e., proportional to its mass accretion rate, \dmvirdt.
\\
\item[(ii)]  The rate at which the gas that was previously ejected due to outflows 
is re-infalling into the galaxy's ISM, \dmfalldt. 
\\
\item[(iii)]  The gas mass that is lost due to star formation 
corrected by the fraction of the material that is instantaneously returned into the ISM, $(1-R)\times\sfr$.
\\
\item[(iv)] The gas mass loss of the galaxy's ISM ejected due to outflows, \dmoutdt. 
\end{description}
Thus, the equation that governs the gas mass growth is 
\begin{equation}
	\mdgasdt = \dmgravdt + \dmfalldt - (1-R) \sfr - \dmoutdt.
\end{equation}
It is more convenient to rewrite \mdgasdt\ as a function of \dmvirdt\ and \sfr. 
To do so, we define \xieff\ as the efficiency with which the inflowing cosmological baryons 
will penetrate down to the galaxy's ISM, \etaout\ as the mass loading factor of gas outflows, and 
\etafall\ as the mass loading factor of gas 
re-infalling. Hence,
\begin{equation}
	\mdgasdt = \fb \xieff \dmvirdt  - ( \etaout  - \etafall +1 - R ) \sfr.
		\label{mgasdot}
\end{equation}

 
 \subsection{Equilibrium Condition}
 
 In the equilibrium solution, galaxies are regulated only between inflow, outflow and
star-formation. The net change of the gas mass within the ISM is zero, $\mdgasdt=0$. 
Within this assumption, the $\sfr$ is given by
\begin{equation}
	\sfr = \frac{\fb \xieff}{\eta  + 1-R} \dmvirdt.
	\label{equil_model}
\end{equation}
Here we define $\eta =  \etaout - \etafall$ as the net mass loading factor. 

At this point, it is worth mentioning why the approach followed in this paper is 
particularly relevant.
This similarity between equations (\ref{SFR_model}) and (\ref{equil_model}) is not a coincidence.  It reflects the fact that a time-independent \shmr\ is {\it compatible with the equilibrium condition}, as
we will show in the next section.

\subsection{Inflows, Outflows and Re-infall}	

\begin{figure*}
	\vspace*{-160pt}
	\hspace*{-10pt}
	\includegraphics[height=8in,width=8in]{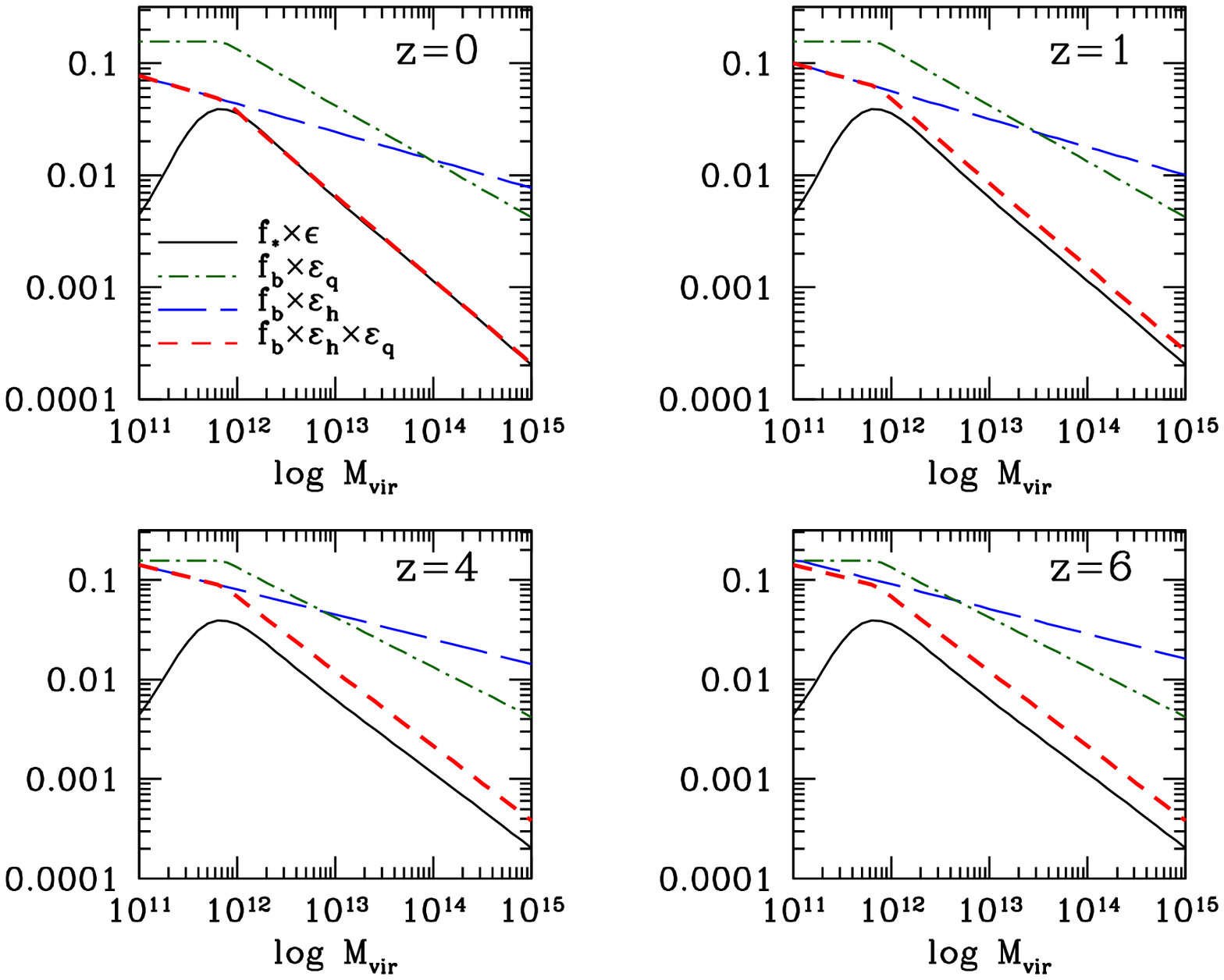}
		\caption{Effective penetration parameter \xieff\ at $z=0,1,4,$ and 6. The vertical axis is
		either, $f_*\times\epsilon$,  $f_{b}\times\cal{E}_{\rm h}$,  $f_{b}\times\cal{E}_{\rm q}$ or $f_{b}\times\xieff$. 
		Here $1-\xieff$ is the
		fraction of gas that never makes it into the galaxy as a result of ``preventive feedbacks'' associated with gas heating by virial shocks ($\cal{E}_{\rm h}$, equation \ref{eps_h})  and AGN feedback ($\cal{E}_{\rm q}$, equation \ref{eps_q}). As explained in the text, $\cal{E}_{\rm h}$ is based on results of other papers.
 	} 	\label{feedback_param}
\end{figure*}

\begin{figure*}
	\vspace*{-360pt}
	\hspace*{-20pt}
	\includegraphics[height=8in,width=8in]{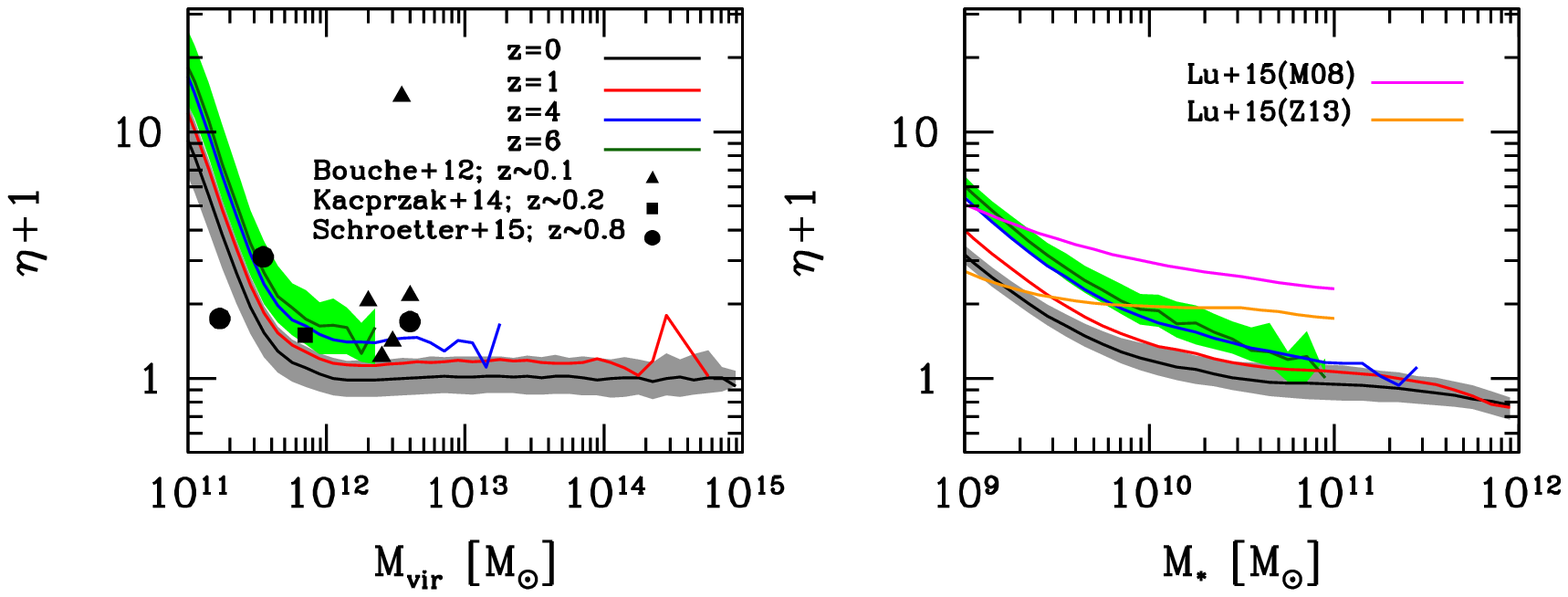}
		\caption{{\bf Left Panel:} Net mass loading factor, $\eta =  \etaout - \etafall$
		as a function of halo mass  at $ z = 0, 1,4,$ and 6, obtained assuming preventive feedback described by ${\cal{E}}_{\rm   eff} = {\cal{E}}_{\rm h} \times {\cal{E}}_{\rm q}$.  The calculated dispersion is shown at $z=0$ and 6.
		We compare our results with observational constraints on the
		mass loading factors from \citet{Bouche+2012} for a sample at $z\sim0.1$, \citet{Kacprzak+2014} for
		a galaxy sample at $z\sim0.2$ and \citet{Schroetter+2015} for galaxies at $z\sim0.8$. Note that these 
		observational constraints are referred to outflowing mass loading factors.
		{\bf Right Panel:} Net mass loading factor and its dispersion as a function of galaxy stellar mass.  Empirical 
		constraints on the mass loading factor based on an analytical model for galaxy 
		metallicity in \citet{Lu+2015} are shown as the magenta
		and orange lines when using gas phase metallicity constraints from \citet{Maiolino+2008} and \citet{Zahid+2013} $z\sim0$, 
		respectively. Similarly above, these results are referred to outflowing mass loading factors. 
		 	}
			\label{loading_fact}
\end{figure*}

Now we combine the equilibrium condition with the time-independent SHARC assumption, and
call the combination E+SHARC. That is, we combine
equations (\ref{SFR_model}) and (\ref{equil_model}) to give
	\begin{equation}
	 	\frac{\fb\xieff}{\eta  + 1-R} = \frac{f_*\epsilon}{1 - R}.
		\label{xieff}
	\end{equation}
It is illuminating	 how the above equation explicitly relates the parameters
from the equilibrium condition (left-hand side) to the SHARC assumption (right-hand side). 
Substantial progress has been made 
in modeling the left hand side of equation (\ref{xieff}), see e.g., 
\citet{Bouche+2010,Dave+2011,Mitra+2015}. Nonetheless 
it is still challenging mainly because it involves many physical processes that are poorly constrained. 
In the time-independent \shmr\ model, however, the value 
of the left-hand side in equation (\ref{xieff}) is constrained by the known value of the right-hand side. While the above does not give any information
of the halo mass and redshift dependence of \xieff\ and $\eta$ separately, this is possible 
if one uses prior information based on models of galaxy formation. 
For example, outflows and re-infall are thought to be more relevant to halos of mass
$M_{\rm vir} \sim 10^{11} M_\odot$ than in halo more massive than 
$M_{\rm vir} \sim 10^{12} M_\odot$ (as we will see later in Figure \ref{loading_fact}). If we assume that such galaxies 
accrete at the maximum rate, i.e., $\xieff\sim1$, a crude estimation for the 
net mass-loading factor is given by $\eta\propto (f_* \epsilon)^{-1}$.  For lower-mass galaxies, preventive
feedback may lead to $\xieff <<1$, as we mention below.
The situation is again different at higher masses, where the accretion of cold gas is diminished and 
outflows and re-infall are less relevant, i.e., $\eta\sim0$. Thus at high mass 
we expect that $f_{\rm b}\xieff = f_*\epsilon$.

The penetration parameter \xieff\ is the result of various forms of preventive feedback, including: 
\begin{description}
\item[(i)] Photoionization heating, $\cal{E}_{\rm ph}$. This term only affects very low-mass 
halos, $\mvir \lesssim 10^{9}\msun$, which we do not consider in this paper.
\\
\item[(ii)]  Heating of the inflowing cosmological baryons due to energetic winds, $\cal{E}_{\rm w}$.  
Winds are more significant in halos lower than $\mvir\sim10^{11}\msun$.
\\
\item[(iii)] Heating of inflowing gas as it crosses virial shocks, $\cal{E}_{\rm h}$. 
This term becomes relevant in halos more massive than $\mvir\sim10^{12}$. 
\\
\item[(iv)]  Any process in massive halos that prevents 
cooling flows from reaching the central galaxy's ISM, for example 
because of maintenance-mode feedback from super-massive black holes ($\cal{E}_{\rm q}$),
which also keeps quenched galaxies quenched.
This term becomes more relevant in halos more massive 
than $\mvir\sim10^{12}$.   
\end{description}
Following \citet{Dave+2012} the resulting \xieff\ is given by
\begin{equation}
\xieff=\cal{E}_{\rm ph}\times  \cal{E}_{\rm w}\times  \cal{E}_{\rm h}\times  \cal{E}_{\rm q},
\end{equation}
For simplicity, we will ignore the impact of 
$\cal{E}_{\rm ph}$ and $\cal{E}_{\rm w}$ -- i.e.,
we assume that $\cal{E}_{\rm ph}\sim\cal{E}_{\rm w}\sim $ 1. 
This is well justified given the halo mass scales $M_{\rm vir} \grtsim 10^{11} M_\odot$ analyzed in this paper. 

We now describe the functional forms we assume for  $\cal{E}_{\rm h}$ and  $\cal{E}_{\rm q}$.
From analysis of hydrodynamic simulations, \citet{Faucher+2011}  derived the
halo mass and redshift dependence of $\cal{E}_{\rm h}$ given by 
\begin{equation}
{\cal{E}}_{\rm h}  ( \mvir,  z) = {\rm min} \left\{ 1, 0.47 \left( \frac{ 1 + z}{ 4 } \right)  \left( \frac{ \mvir } { 10 ^ {12} \msun } \right)^{-0.25}\right\}
\label{eps_h}
\end{equation}
\citep[see also][]{Dave+2011,Dave+2012}.
Figure \ref{feedback_param} shows the feedback parameter $\cal{E}_{\rm h}$ multiplied by
the universal baryon fraction, $\fb$, as the 
blue long dashed-line at $z = 0, 1, 4$ and $z = 6$. In the same figure, we show the halo 
mass dependence of  $f_* \times\epsilon$ as the black solid line. 
Recall that we assume that $f_* \times\epsilon$ is independent of redshift. 

If $f_*\times\epsilon =  f_{\rm b}\times \cal{E}_{\rm eff} $, then the mass loading factor $\eta=0$. 
The difference between these two quantities is therefore related to $\eta$. 
The point of maximum approach between these two curves is when $f_* \times\epsilon$
reaches its peak value at Milky Way sized halos, $\mvir\sim10^{12}\msun$, 
where $\fb\times\cal{E}_{\rm h}$ is a factor of $\sim1.2$ higher than $f_* \times\epsilon$
at $z=0$. At $z=6$ the situation is qualitatively different and 
$\fb\times\cal{E}_{\rm h}$ is a factor of $\sim2.5$ higher than $f_* \times\epsilon$ in $\mvir\sim10^{12}\msun$ halos. Then the mass loading factor should increase at high redshift. 

Halo mass quenching is more relevant for high mass halos.
This imposes the constraint that any functional form proposed for $\cal{E}_{\rm q}$ should reproduce
the fall off at higher masses of the term $f_* \epsilon$. Given the uncertain redshift dependence of
$\cal{E}_{\rm q}$, we will assume for simplicity that it is independent of redshift. 
The functional form $ \cal{E}_{\rm q}$ that 
describes the fall-off of $f_* \epsilon$ at $z=0$ is given by
\begin{equation}
{\cal{E}}_{\rm q} (\mvir) = {\rm min} \left\{ 1, 0.85 \left( \frac{ \mvir } { 10 ^ {12} \msun } \right)^{-0.5}\right\}.
\label{eps_q}
\end{equation}
Note that at $z = 0$ for halos more massive than $\sim10^{12}\msun$, $\xieff\sim \epsilon\times f_*/\fb$.
Such a fall-off  is thus necessary in order to make SHMR+equilibrium assumptions
work, in other words, equation (\ref{equil_model}).
The green long dashed-dotted lines in Figure~\ref{feedback_param} show $\cal{E}_{\rm q}$.
At higher redshifts $\xieff>\epsilon\times f_*/\fb$ implying that the mass-loading factor becomes
more important at high redshifts in high mass galaxies.

Next,  in equation (\ref{xieff}) we use the functional forms described in equations 
(\ref{eps_h}) and (\ref{eps_q}) to deduce a relation for the net mass loading factor: 
	\begin{equation}
		\eta = \left[ \frac{\fb}{f_* (\mvir)}\frac{\xieff( \mvir,  z)}{\epsilon (\mvir)} - 1 \right] (1 - R).
	\end{equation}
The left hand panel of Figure \ref{loading_fact} shows the net mass loading factor, $\eta =  \etaout - \etafall$,
as a function of halo mass at $z = 0, 1, 4$ and $6$. Note that the generic redshift evolution of 
$\eta$ is governed by the evolution of \xieff.  For halos less massive than $\sim10^{11.5}\msun$, Figure~\ref{loading_fact} shows that 
the mass loading factor approximately scales as a power law with a power that is roughly 
independent of redshift, $\eta\propto M_{\rm vir} ^ {-2.13}$. Equivalently, we find that for galaxies with stellar mass below $\sim10^{9.7}\msun$ the mass loading factor scales as 
$\eta\propto M_{\rm *} ^ {-1.07}$.  Mass loading factors are predicted to be very small for halos more massive
than $\sim10^{12}\msun$, especially at low redshifts. For comparison we include 
observational constraints on the mass loading factors from \citet{Bouche+2012} 
for a sample at $z\sim0.1$, \citet{Kacprzak+2014} for
a galaxy sample at $z\sim0.2$ and \citet{Schroetter+2015} for galaxies at $z\sim0.8$.
Empirical constraints on the mass loading factor based on an analytical model for galaxy 
metallicity in \citet{Lu+2015} are shown as the magenta
and orange lines when using gas phase metallicity constraints from \citet{Maiolino+2008} and \citet{Zahid+2013} at
$z\sim0$, 
respectively. Note that these comparisons are referred to outflowing mass loading factors. Nevertheless,
at lower masses, this comparison is fair since most of the outflowing gas is the most relevant contribution to the net mass loading factor.

In this Section we presented a simple framework that clarifies how the net mass loading factor
is connected to preventive feedback in the context of the equilibrium time-independent \shmr\ model. 
As long as the SFR is driven by MAR these assumptions can be generalized in 
the same framework, as we mention briefly in the discussion section.

\section{Specific Star Formation Rates from SHARC}
\label{SFR_section}

\subsection{SHARC Compared with Observations}

\begin{figure*}
	\vspace*{-20pt}
	\hspace*{0pt}
	\includegraphics[height=6in,width=6in]{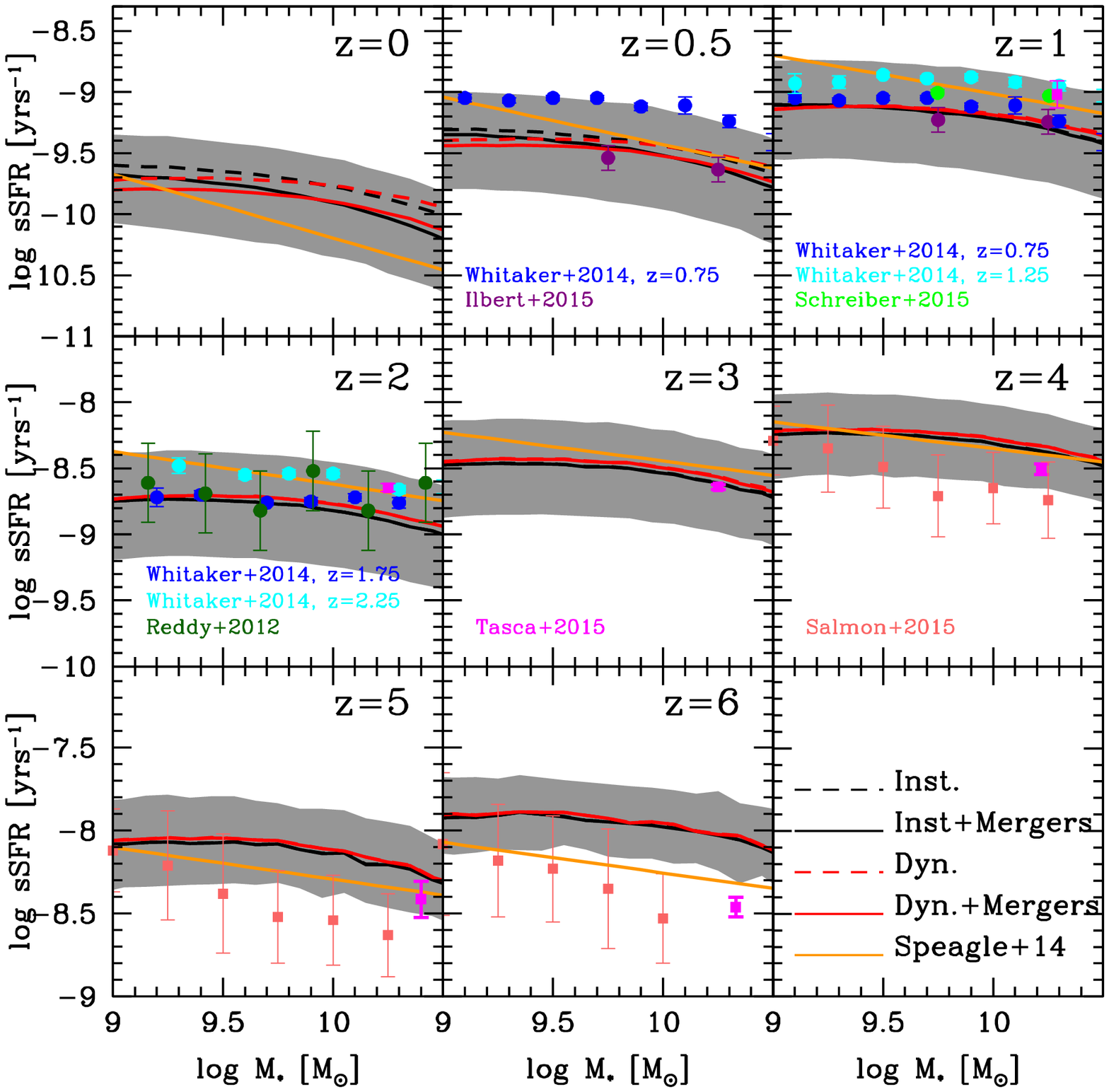}
		\caption{ Redshift evolution of s$\sfr$s derived in the time-independent \shmr\ model (SHARC assumption). 
		The red and black curves are the medians of the s$\sfr$s from the dynamically-time-averaged and
		instantaneous mass accretion rates, respectively, the dispersion about the median
		sSFRs with the gray band calculated from instantaneous mass accretion rates. .  
		These are corrected for mergers (see \S\ref{Mergers}) while the
		corresponding dashed curves are not.  These are compared with the \citet{Speagle+2014}
		summary of observed s$\sfr$s on the main sequence (orange curve), and also with 
		other recent measurements.  Note that the \citet{Whitaker+2014} measurements from
		$z=0.5-1.0$, listed as Whitaker+14, are shown in both $z=0.5$ and $z=1$ panels.  In addition, we  		include star-forming data from \citet{Ilbert+2015,Salmon+2015,Schreiber+2015,Tasca+2015} and \citet{Reddy+2012}.
		}
	\label{sfr_evol}
\end{figure*}

\begin{figure*}
	\vspace*{-130pt}
	\hspace*{40pt}
	\includegraphics[height=6in,width=6in]{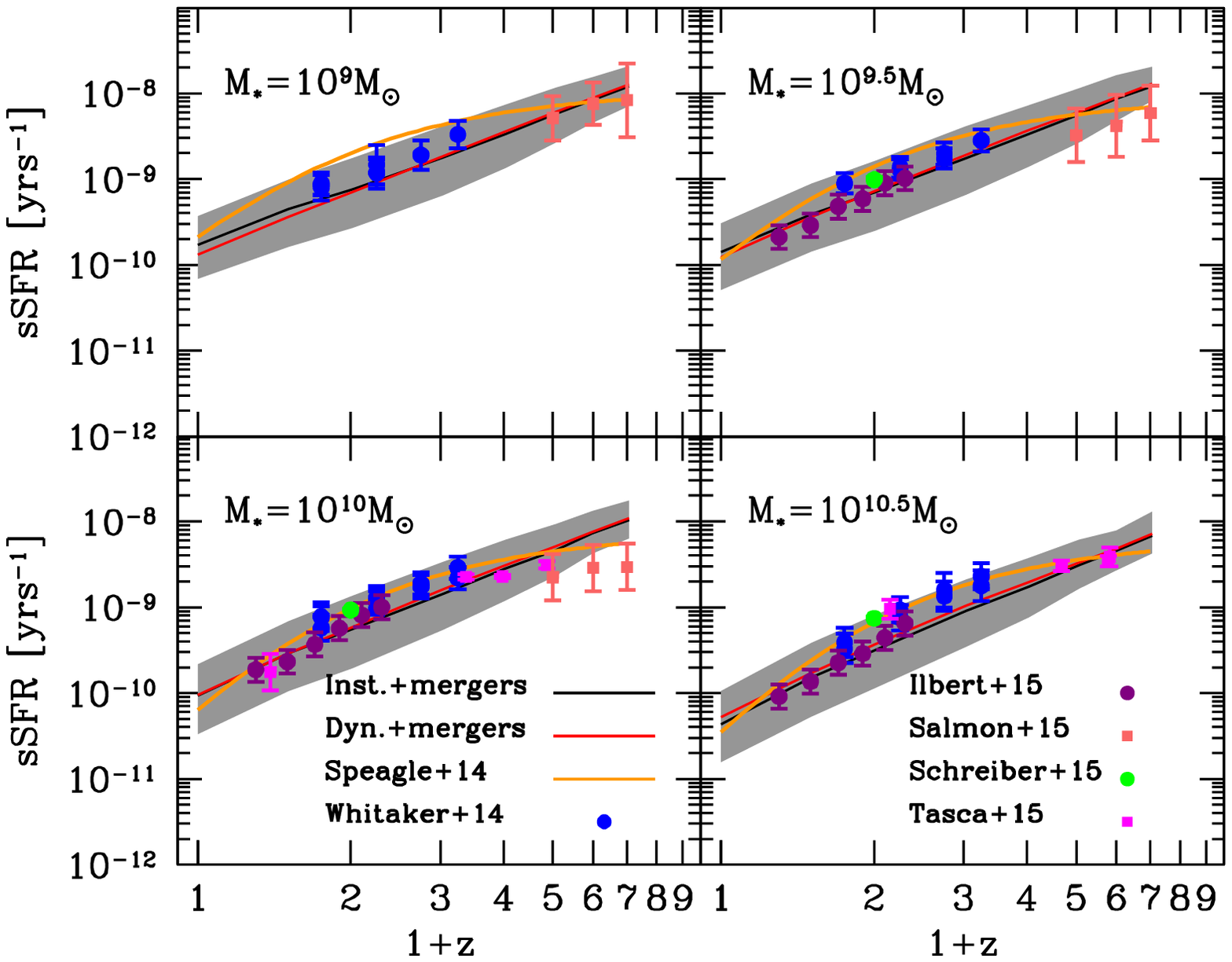}
		\caption{Specific star formation rates as a function of redshift $z$ for stellar masses $\ms= 10^{9},  
		10^{9.5}, 10^{10}$ and $10^{10.5}\msun$ from time-independent \shmr\ model.  
		The red and black curves are the s$\sfr$s, from both dynamically-time-averaged and
		instantaneous mass accretion rates, respectively, with the gray band representing 
		the dispersion in the latter.  Both are corrected for mergers.  The orange curve is the
		 \citet{Speagle+2014} summary of observed sSFRs on the main sequence.  Observations
		 from \citet{Whitaker+2014}, \citet{Ilbert+2015}, \citet{Salmon+2015}, 
		\citet{Schreiber+2015} and \citet{Tasca+2015} are also included.
 	}
		\label{sSFR}
\end{figure*}

We have now collected together all the tools needed to follow several aspects of galaxy evolution
while galaxy stellar masses are in the range $\ms = 10^{9}\msun$ to $10^{10.5}\msun$. 
We start by showing the evolution in the slope and zero-point of the
star-forming main sequence inferred by the time-independent \shmr\ (SHARC model) in Figure \ref{sfr_evol}. 
Recall that when assuming a time-independent \shmr, stellar mass growth can be inferred directly
from halo mass accretion rates via $\dmsdt =  f_* \times \epsilon \times \dmvirdt$, with the corresponding
$\sfr=\dmsdt/(1-R)$.
Black solid lines show
results using instantaneous mass accretion rates, \dmvirdt, in equation (\ref{dmsdt}). Red 
solid lines show the $\sfr$s when using mass accretion rates smoothed over a dynamical
time scale, \dmvirdyn, instead. The gray band indicates the intrinsic scatter around the
star-forming main sequence when using \dmvirdt. Note that our model sSFRs were corrected in order
to take into account the contribution of mergers to stellar mass growth, 
as explained in \S\ref{Mergers}. We show the 
resulting sSFRs without this merger correction with the black and red 
dashed lines when using \dmvirdt\ and \dmvirdyn\ respectively. Note that the contribution
from mergers becomes more important for redshifts $z < 0.5$. Hereafter, we will focus
our discussion on the merger-corrected results, also shown as the solid lines in Figure~\ref{sSFR}.

Both \dmvirdt\ and \dmvirdyn\ produce similar relations at all redshifts. 
This is expected due to the similarities shown between \dmvirdt\ and \dmvirdyn\ in Figure~\ref{HMAR}. 
We note that the resulting slope of model the star-forming main sequence when using 
\dmvirdt\ at $z=0$ is $0.73$ and increases as a function of redshift to a value of 0.87 at $z = 6$.
(Similar slopes are derived for \dmvirdyn .)
This is consistent with observed slopes derived from SDSS galaxies
\citep[see e.g,][]{Elbaz+2007, Zahid+2012,Salim+2007} as well as from high-redshift galaxies
\citep[see e.g,][]{Santini+2009,Karim+2011,Reddy+2012}. 

In Figure \ref{sfr_evol}, we reproduce the best fit reported in \citet{Speagle+2014} 
to the star formation main sequence as the orange curve, as well as more 
recent observations. 
\citet{Speagle+2014} used a compilation of 25 observations from
the literature to study the star formation main sequence from $z=0$ to 6. The authors carefully
calibrated observational $\sfr$s, correcting for different 
assumptions regarding the stellar IMF, \sfr\ indicators, 
SPS models, dust extinction, emission lines and cosmology, among 
the most important calibrations. Hence, their best-fitting model represents a
robust inference of the redshift evolution of the star-forming main sequence. 
Our derived star-forming main sequences are in good agreement with \citet{Speagle+2014}
and within the $1\sigma$ intrinsic scatter of our relations at almost all redshifts. Note,
however, that there are some systematic deviations between our model predictions and 
the observations.  While these differences could be due to redshift-dependent systematic biases in
the observationally-inferred s$\sfr$s, it is interesting to discuss these differences in the light
of the constant SHMR model. 

First, the observed $\ssfr$s of galaxies at $ z > 4$ are systematically lower than the time independent 
SHMR model predictions. 
These differences increase at $z=6$. The disagreement between the constant SHMR predicted SFRs
and the observations implies that the changing SHMR must be used, as in equation (\ref{full_dmsdt}),
at least at high redshift.

Between $z=4$ and $z=3$ the observed star-forming sequence is consistent with the 
SHARC predictions. Between $z=2$ and $z=0.5$, the observed $\ssfr$s are slightly above the 
SHARC predictions. This departure occurs at
the time of the peak value of the cosmic star formation rate. 

After the compilation carried out by
\citet{Speagle+2014}, new determinations of the sSFR have been published, particularly
for redshifts $z < 2.5$. In Figures \ref{sfr_evol} and \ref{sSFR}, we reproduce new data published in
\citet{Whitaker+2014,Ilbert+2015,Salmon+2015,Schreiber+2015} and \citet{Tasca+2015}. This new set of data
agrees better with our model between $z=2$ and $z=0.5$, implying that the time-independent 
SHMR (SHARC assumption) may be nearly valid across the wide redshift range  from $z\sim4$ to $z\sim0$, 
a remarkable result.  However, it is not
clear whether this is valid since the newer observations have not been recalibrated as in \citet{Speagle+2014}.

\begin{figure}
	\vspace*{-100pt}
	\hspace*{-20pt}
	\includegraphics[height=4in,width=4in]{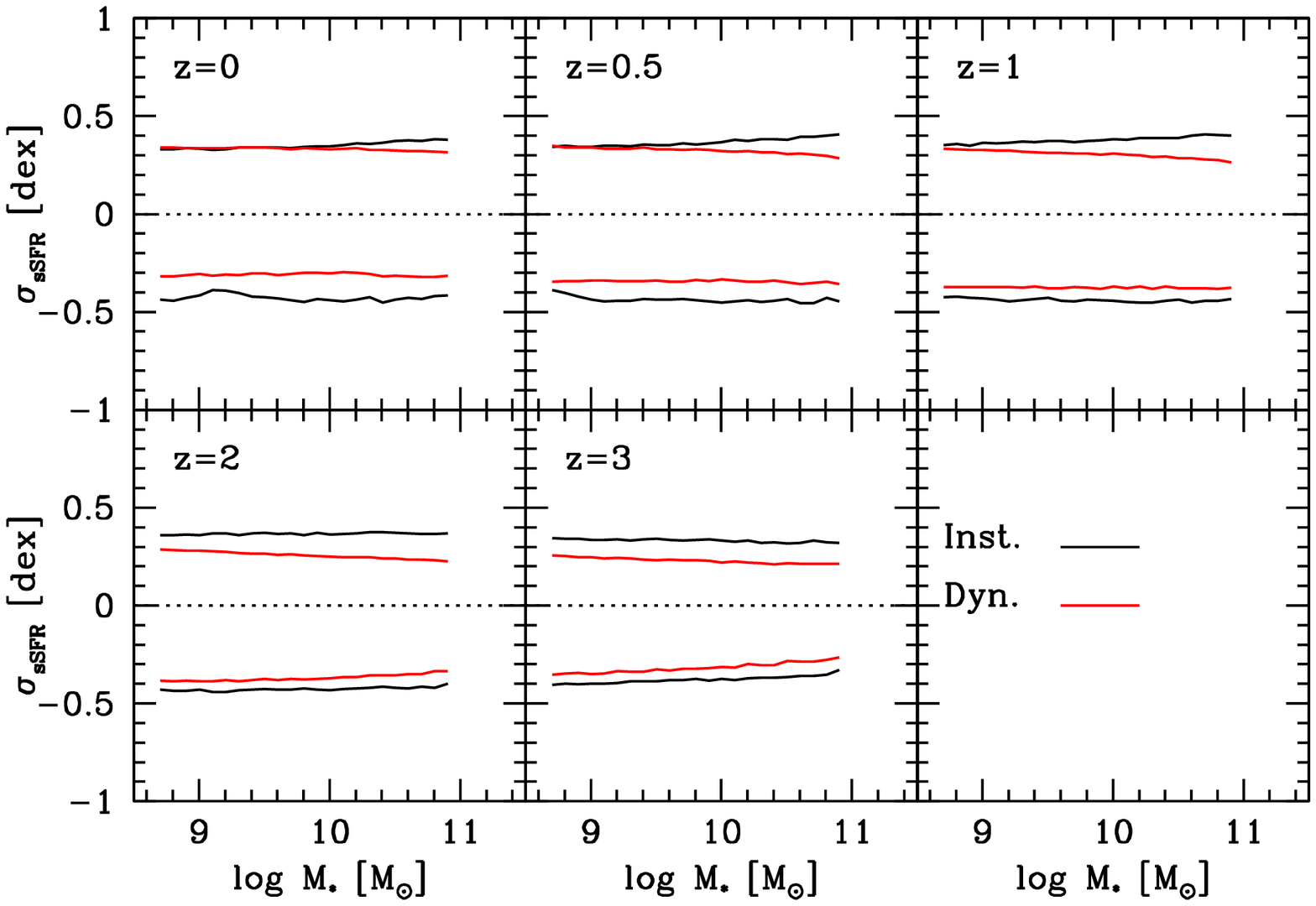}
		\caption{ Scatter of the s$\sfr$ for main-sequence galaxies predicted in our model.
 	}
	\label{sSFR_scatter}
\end{figure}

\subsection{Scatter of the s\sfr\ Main Sequence}

\begin{figure*}
	\vspace*{-500pt}
	\hspace*{0pt}
	\includegraphics[height=12in,width=12in]{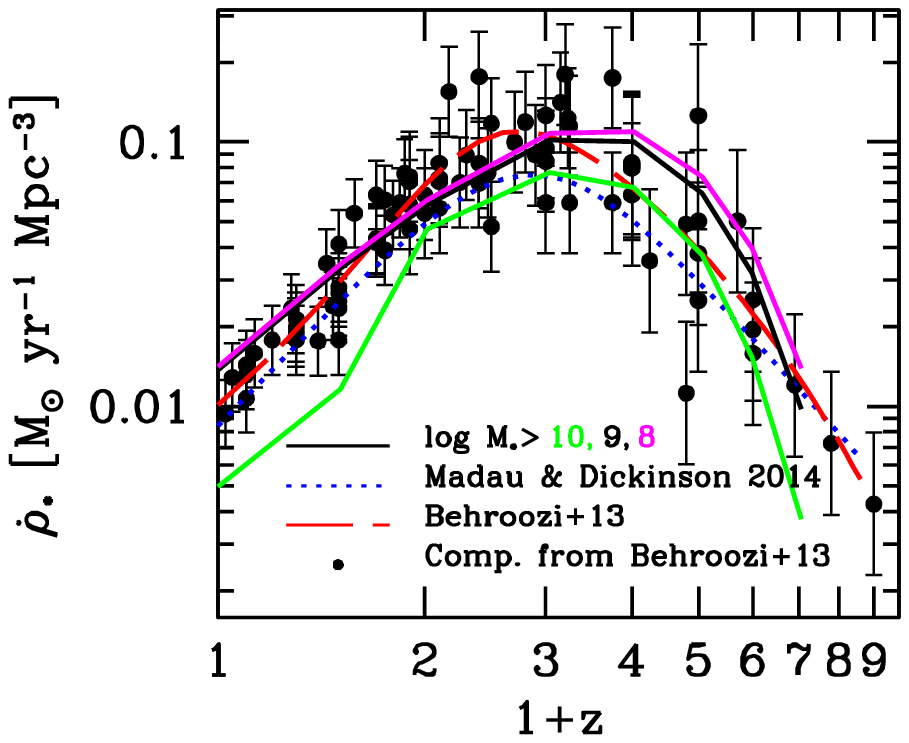}
	\caption{
	Cosmic star-formation rate (CSFR) as a function of $z$, for both our model (with
	results integrating down to $\log M_*/M_\odot =$ 8, 9, and 10 shown as
	magenta, black, and green curves), and for fits by \citet{Behroozi+2013a}
	(shown as a long-dashed red curve), \citet{Madau+2014} (dotted blue curve),
	and sets of compiled observations \citep{Behroozi+2013a}. As discussed in the 
		text the predicted CSFR in the SHARCK are probably related to the failure of the time-independent  
		assumption. 
 	}
	\label{CSFR}
\end{figure*}

We now turn our discussion to the scatter of the star-forming 
main sequence, displayed in Figure \ref{sSFR_scatter}. When using \dmvirdt, the scatter is nearly 
independent of redshift and it increases very slowly with mass for 
$z < 2$. The value of the scatter is not symmetric and has values of 
$\sigma \sim 0.35 - 0.45$ dex. In contrast, for $z  \grtsim 2$ the scatter decreases 
with mass. Instead using \dmvirdyn\ produces a scatter
more symmetric and practically independent of mass and redshift
below $z=2$. The scatter takes a value of $\sigma \sim 0.35$ dex. At
high redshifts, the scatter decreases with increasing mass.
We cannot make a direct 
comparison with observations due the uncertainties affecting 
the measurements of both stellar masses and $\sfr$s. Nevertheless,
attempts to deconvolve the intrinsic scatter from measurement errors,
particularly for SDSS galaxies, have found that the star-forming 
main sequence has a scatter of $\sigma \sim 0.3$ dex
\citep[see, e.g.,][]{Salim+2007,Speagle+2014,Schreiber+2015}. 
New results on star-formation rates from CANDELS optical-IR colors also support a main-sequence scatter of $\sim$0.3 dex that is remarkably uniform from $z = 0.5$ to 2.5 at all masses 
$10^9\msun$ and above (Fang et al., in prep.). The behavior of the SFR dispersion can perhaps
be understood as reflecting the Central Value Theorem \citep{Kelson2014} applied to
the halo MAR.

In this paper, we are using the MARs for all distinct dark matter halos
to predict the SFR on the galactic main sequence and its scatter.
What if we instead only used the halos that host star-forming central
galaxies?  In work in progress, we have found that doing this at $z\sim 0$ using age 
matching \citep{HearinWatson2013} and similar methods
results in a somewhat smaller scatter in the predicted sSFR of about 
0.3 dex, in even better agreement with observations.  
It is not clear whether this will also be true for $z>0$, however. 

\begin{figure*}
	\vspace*{-500pt}
	\hspace*{0pt}
	\includegraphics[height=12in,width=12in]{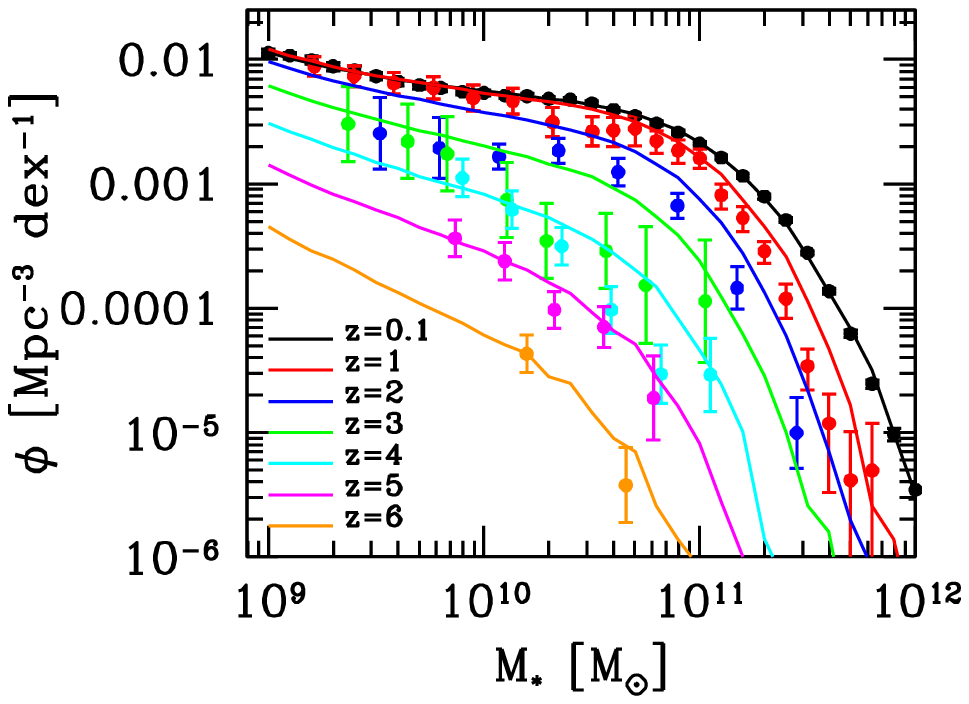}
		\caption{Redshift evolution of the galaxy stellar mass function (GSMF) as predicted by the 
		time-independent \shmr\ (SHARC assumption), and as observationally-derived in \citet{RP+2015} at $z = 0.1$; \citet{Moustakas+2013} and \citet{PG+2008} at $z = 1$; \citet{Marchesini+2009} at $z = 2$; \citet{Mortlock+2011} at $ z = 3$; \citet{Lee+2012} at $z = 4$ and 5; and \citet{Stark+2013} at $z = 6$.  	
		}
	 \label{fig_GSMF}
\end{figure*}

\section{SHARC Evolution of the Cosmic Star Formation Rate \&\ Stellar Mass Function}
\label{CSFR_SMF}

The SHARC model can be used to calculate the total cosmic star formation rate (CSFR) as a function of time.  
This is shown in Figure \ref{CSFR}, which plots the results of using \dmvirdt\ but similar results are obtained if \dmvirdyn\ is used instead. For comparison, we also reproduce a compilation
presented in \citet{Behroozi+2013a}, including data from UV, UV+IR, IR, 
H$\alpha$ and 1.4 GHz, as well as a recent fit to observations including
UV and IR by \citet{Madau+2014}. The peak of our CSFR occurs at $z\sim3$ 
which is earlier than the fit of \citet{Madau+2014} and the data compiled in  
\citet{Behroozi+2013a}, peaking at $z\sim2$. 

Figure \ref{CSFR} shows that the CSFR from the SHARC model is 
higher than most of the observations at $z>2$, although as we showed in the previous section, 
the SFR predicted by the model agrees between z=3 and 4 with \citet{Speagle+2014}. However, at
$z>4$, where the model predicts high SFRs, this is a failure of the constant SHMR condition. 

Figure \ref{fig_GSMF} shows the redshift evolution of the \gsmf\ calculated from the 
time-independent \shmr, SHARC assumption. 
In the same Figure we compare to some observational inferences as indicated in the caption. 
The predicted evolution of the \gsmf\ is consistent with observational inferences at different redshifts.  

\section{SHARC+Equilibrium Bathtub: Metallicities}
\label{Metals_EQC}
\begin{figure*}
	\vspace*{-160pt}
	\hspace*{0pt}
	\includegraphics[height=6.5in,width=6.5in]{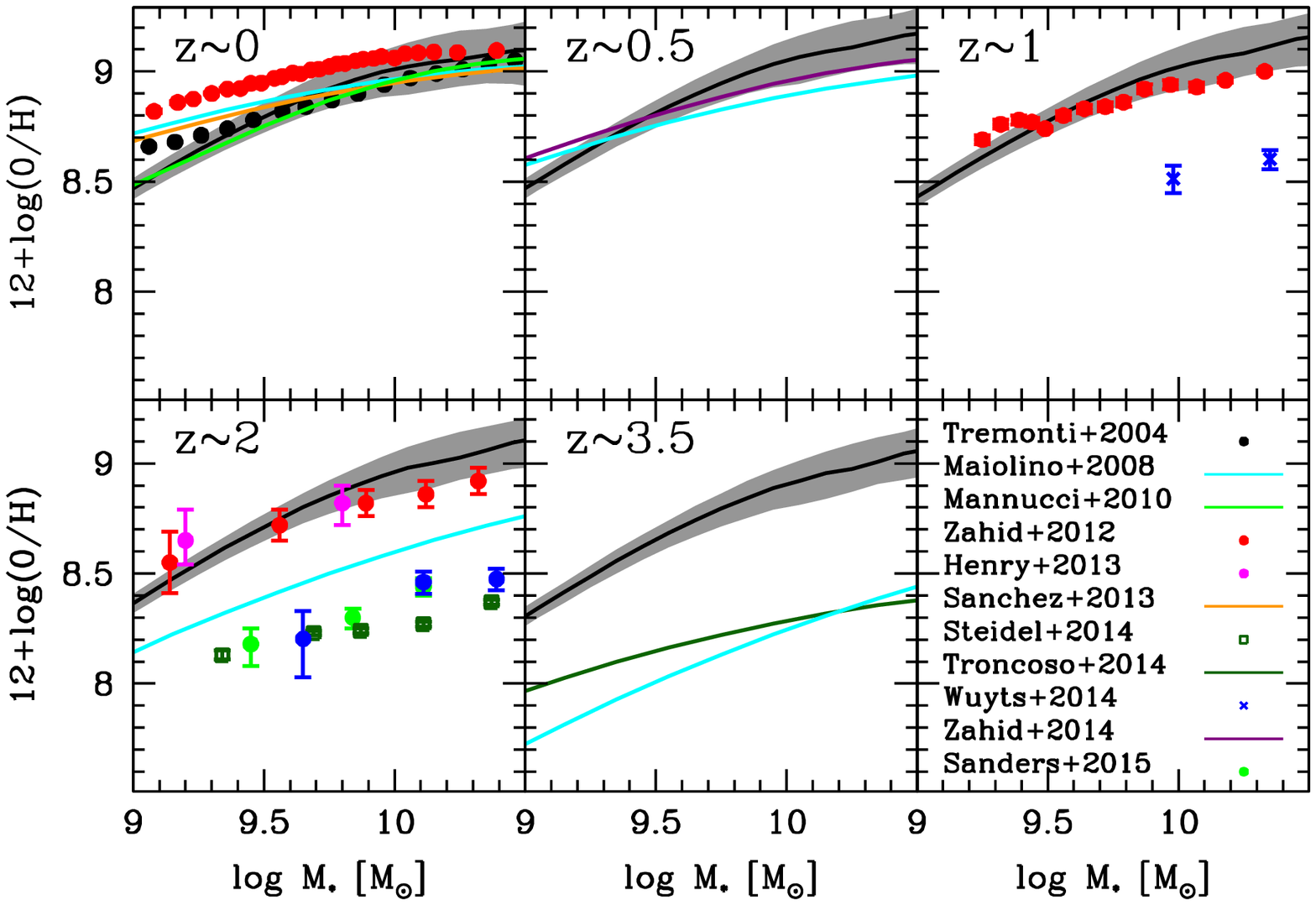}
		\caption{Metallicities as a function of stellar mass, both as predicted by the E+SHARC model  and as observed by \citet{Tremonti+2004,Maiolino+2008,Mannucci+2010,Zahid+2012,Henry+2013,Sanchez+2013,Steidel+2014,Troncoso+2014,Wuyts+2014,Zahid+2014} and \citet{Sanders+2015}.
		Conversion from \zism\ to 12+log(O/H) following \citet{Feldmann2015}. 
		The E+SHARC model predictions are in good agreement with observations below
		$z < 2$ but disagree at higher redshifts probably reflecting the failure of the 
		equilibrium condition. 
 	}
	\label{metallicities}
\end{figure*}

In \S4 and \S5, we only used the time-independent SHMR (SHARC). 
Now we return to the bathtub model \S3.
Using a time-independent SHMR plus the Equilibrium assumption, 
E+SHARC, we can predict the metallicities in the ISM. Metallicity is defined as $\zism\equiv\mz/\mgas$, where $\mz$
is the mass of metals in the gas phase within the ISM and \mgas\ the total cold gas mass.
The change of the metal mass within the ISM of a galaxy is given by
\begin{eqnarray}
	\mzdt = y\times\sfr +  \zigm\fb \xieff \dmvirdt +&  &\nonumber \\ (\zfall\etafall - (1-R) \zism - \zout\etaout)\sfr.
	\label{metal_mass}
\end{eqnarray}
The yield $y$ in equation (\ref{metal_mass}) is the metal mass  in the gas phase formed 
and returned to the ISM per unit \sfr. We assume that the metal yield is instantaneous,
and use $y/(1-R) = 0.054$ as derived in \citet{Krumholz+2012} for solar metallicites. 
The terms \zigm, \zfall, and \zout\ are the metallicities of the intergalactic medium, the
re-infall of previously ejected material, and the inflows of the ISM, respectively. 

 Let $\alpha_r$ and $\alpha_w$ be defined as $\zfall = \alpha_r\zism$ and $\zout = \alpha_w\zism$.
 Also, let the metallicity of inflowing material from the 
intergalactic medium be some fraction $\alpha_{\rm IGM}$ of the galaxy's ISM 
($\zigm = \alpha_{\rm IGM}\zism$).
If we use the fact that $\zism=\mz/\mgas$ in equation (\ref{metal_mass}),  it follows that
the metallicity in the ISM is
\begin{eqnarray}
	\zism =  \hspace{2.6in} & &\nonumber \\
	\frac{y\times\sfr - \dzismdt\times\mgas}{(\alpha_w\etaout -  \alpha_r\etafall + 1-R)\sfr -   \alpha_{\rm IGM} \fb \xieff \dmvirdt}.
\end{eqnarray}

In what follows we will assume that the outflowing metallicity is equal to the ISM 
metallicity ($\alpha_w\sim1$). Also, we will assume that the galaxy ISM metallicity
changes slowly compared to the re-infall time, so that 
$\alpha_r\sim1$, and that the metallicity in the IGM is close to zero, i.e., $\alpha_{\rm IGM} \sim0$. 
If the galaxy's metallicity
evolves only slowly with time ($\dzismdt\sim0$), the
above equation can then be re-written as
\begin{equation}
	\zism = \frac{y}{\eta  + 1-R}.
\end{equation}
This familiar equation is similar to that in \citet{Dave+2012}. 
Under these assumptions the metallicity of the ISM is controlled by the
net mass-loading factor $\eta$, which is itself controlled by the 
efficiency $\xieff$  (see equation \ref{xieff}). 
In other words, the evolution of the metallicity in a galaxy's ISM is
driven only by the efficiency at which the baryons penetrate down to the galaxy.
Note that this is no longer valid if enriched outflows are considered, i.e., $\alpha_w > 1$. 
\citet{Feldmann2015} studied the general case and concluded that even if galaxies 
are not in strict equilibrium  the outflowing metallicity is close to that of the galaxy's ISM.
Using equation (\ref{xieff}), 
we can thus write \zism\ as
\begin{equation}
	\zism = \frac{f_*\times\epsilon}{\fb\times\xieff} \frac{y } {1 - R}.
\end{equation}

Note that a direct consequence of the 
equilibrium condition is that the scatter in $\zism$ is simply a
consequence of the scatter in the SHMR, i.e.,
that halos of the same mass have a range of values of $f_*$. 
This is an important conclusion because if the observed scatter in \zism\ is similar to the
intrinsic scatter of the SHMR, that would provide further evidence for the equilibrium condition. 

The resulting metallicities are compared with observations in Figure~\ref{metallicities}. Agreement would support the 
E+SHARC assumption, our preventive feedback assumptions, and our simple metallicity treatment.
The model predictions
are actually in good agreement with some of the observations from $z=0$ to $\sim 2$, but it is hard to draw strong conclusions because of the
disagreements between different observations.  However it is clear that our constant SHMR equilibrium model predicts metallicities that are much higher than observed at $z\sim 3.5$, which is a consequence of the model's overprediction of the SFR at high redshift.

\section{Conclusions}
\label{Conclusions}

\subsection{Summary}

In the present paper 
we have investigated to what extent the mass accretion rate of the host halo controls the rate of star formation of
galaxies on the main sequence of star formation.  We were motivated by the realization that the halo mass accretion 
rate (MAR) and its scatter, shown in Figures~\ref{HMAR} and \ref{HMARscatter}, bear a remarkable resemblance to the star 
formation rate on the main sequence and its scatter.  In order to connect these phenomena, we 
have considered an extremely simple -- no doubt oversimplified -- approach in which we made the crucial
assumption that the stellar-to-halo mass relation (SHMR) for central
galaxies in dark matter halos, deduced from SDSS observations
and shown in Figure~\ref{f3}, remains valid at all redshifts. We called this Stellar-Halo Accretion
Rate Coevolution (SHARC) assumption. 
We showed in \S2 that the SHARC assumption allows derivation of galaxy SFRs from halo MARs. This robust 
new result can also be expressed as equation (\ref{virial_SF}) for the ``virial star formation efficiency", 
which only depends on halo mass, \mvir.  In \S\ref{EQC} we showed that 
 the SHARC assumption, i.e., a time-independent SHMR, is compatible with the 
equilibrium condition that determines the amount of gas reaching the interstellar medium due to 
preventive feedbacks to the net mass-loading factor. We call this E+SHARC.  
Assuming reasonable preventive feedbacks based on simulations allows calculation of the mass loading factors and their dispersions shown in Figure~\ref{loading_fact}.

The specific star formation rate (sSFR) and its 
dispersion predicted by the SHARC assumption are shown in Figures~\ref{sfr_evol}, \ref{sSFR}, and \ref{sSFR_scatter}.  Despite the simplicity of our
assumptions, the resulting predictions are in rather good agreement with observations from $z=0$ out
to $z\sim 4$, especially if the most recent observations are used.  At $z\grtsim 4$ the predicted sSFRs are systematically
higher than observations, implying that the redshift-independent SHMR assumption  breaks down at $z \grtsim 4$.

Figures~\ref{CSFR} and \ref{fig_GSMF} show that the cosmic star formation rate density (CSFR) and galaxy stellar mass function predicted by the SHARC assumption is in good agreement with observations at $z \lesssim 1$, but the CSFR 
density is higher than most observations at higher redshifts.  This disagreement at higher redshifts
again arises  because the redshift-independent SHMR assumption is invalid at higher redshifts.

Now, assuming SHARC plus an equilibrium bathtub model (E+SHARC)  and  
 a few additional plausible assumptions regarding preventive feedback and inflow and outflow metallicites, the model also predicts ISM metallicities as a function of galaxy stellar mass and redshift.  These predictions are compared with observations in Figure~\ref{metallicities}. The predictions are in good agreement with  at least some of observations $z\lesssim2$, although the scatter in the data is rather large.  
 At redshift $z\sim3.5$ the predicted metallicities are much higher than observed, indicating that 
 the combination of the Equilibrium condition and SHARC assumption is invalid at high redshift.

\subsection{Implications of SFR Determined by Halo Mass Accretion}

In this section, we look ahead to some important implications that the toy model has for understanding other current modeling techniques.  For example, 
abundance matching, based on galaxy stellar mass and halo mass (or a related quantity such as peak circular velocity) leads to the simplest models relating galaxies to their host dark matter halos.  As mentioned in \S1, such models predict galaxy correlation functions in good agreement with observations both nearby and out to high redshifts.  
But the clustering and galaxy content of dark matter halos are known to be a function of more than just the mass or circular velocity of the halos.\footnote{From the earliest papers on cold dark matter \citep{BFPR,Faber84,Primack84} it was clear that dark matter halos would be characterized by a second parameter beyond mass such as overdensity, which is related to formation time.}
 In particular, the formation time and concentration of halos appear to play a major role.  Halos of much lower mass than the typical mass $\mathcal{M}^\ast(z)$ collapsing at a given redshift $z$ are much more strongly clustered if they formed at high redshift than similar-mass halos that formed at low redshift \citep{Gao+2005}, a phenomenon known as ``assembly bias" \citep[e.g.,][]{MvdBW}.  Halo concentration, $c_{\rm vir} = R_{\rm vir} / R_{\rm s}$ (where $R_{\rm vir}$ is the virial radius and $R_{\rm s}$ is the NFW \citep{NFW} scale radius), is related to halo formation time, with halos of higher concentration at fixed mass forming earlier \citep{Bullock+2001,Wechsler+2002}, and with higher-$c_{\rm vir}$ halos with mass $M_{\rm vir} \ll \mathcal{M}^\ast$ being much more clustered than average \citep{Wechsler+2006}.  \citet{Bullock+2001} had suggested that a natural association is that high-$c_{\rm vir}$ halos host old, red galaxies, and lower $c_{\rm vir}$ halos host young, blue galaxies.  This idea was rediscovered by \citet[][and subsequent papers]{HearinWatson2013}, who showed that filling halos according to this prescription correctly predicts the observed clustering of red and blue galaxies in the SDSS.  It seems surprising that such a simple prescription should work so well.

More recently, \citet{HearinBehroozivdB} showed that the mass accretion rate of dark matter halos at low redshift shows a signal very much like the observed two-halo ``galaxy conformity" \citep{Kauffmann+2013,HearinWatsonvdB+2014}, namely that quenched central galaxies tend to lie in quenched regions as large as 4 Mpc.  A natural explanation for this finding would be that the star formation rate in central galaxies is closely connected with the mass accretion rate of their host halos.  

Figures~\ref{sfr_evol} and \ref{sSFR} show that the sSFR predicted by the SHARC assumption, in which the SFR is proportional to the host halo mass accretion rate, is in rather good agreement with the observations, especially at $z \sim 0$. In more elaborate models of galaxy formation and evolution such as semi-analytic models (SAMs) and hydrodynamic simulations, the star formation rate, morphology, and other galaxy properties are assumed to result from a complex interplay between gas inflows and outflows regulated by stellar and AGN feedback and other processes, involving much recycled rather than recently accreted gas.  
If star-forming galaxies at low redshifts are indeed nearly in equilibrium, then the SFR will in fact be driven by halo mass accretion, which may represent the net result of complex processes considered in more detailed galaxy formation models.
A close connection between halo accretion and star formation may help to explain the success of age matching \citep{HearinWatson2013} and the agreement between halo MAR conformity and galaxy conformity observations \citep{HearinBehroozivdB}.

\subsection{Outlook}

Several modifications can add realism to the simplified model considered here: 
\begin{itemize}
\item Instead of assuming that the stellar-to-halo mass relation (SHMR) is redshift independent, use the evolving SHMR implied by abundance matching to connect halo MAH to galaxy SFR, using
equation (\ref{full_dmsdt}).

\item Instead of assuming that a galaxy is always in equilibrium, assume alternatively that the gas mass grows from high redshifts down to $z\sim 4$ -- i.e., the bathtub fills in the early universe. What early universe combinations of galactic gas mass growth and evolving SHMR predict SFRs and metallicities in agreement with the rapidly improving observations?

\item Explore how changing the assumptions regarding gas penetration efficiency $\cal{E}_{\rm  eff}$  leads to different dependance on halo mass and redshift of mass-loading factors and metallicity growth. 

\item Instead of assuming for simplicity that gas in outgoing winds has the ISM metallicity and that freshly accreted gas has zero metallicity, compare predictions from modified assumptions with improving data on galactic gas metallicity at various redshifts.

\item With the equilibrium condition $\mdgasdt = 0$, the gas
depletion time scale $t_{\rm depl} = M_{\rm gas}/\sfr$ is just proportional to $\sfr^{-1}$,
which implies that the slope of the $t_{\rm depl}$ to $\ssfr$ relation is $-1$. 
But recent papers \citep{Sargent+2014,Huang+2014, Huang+2015,Genzel+2015} find that the slope of the $t_{\rm depl}$ to $\ssfr$ relation is roughly $-0.5$ for main sequence galaxies
at $z=0$ to 3.  This suggests relaxing the equilibrium condition to treat excursions about the main sequence.   

\item In this paper we considered star formation of central galaxies in halos of mass $\sim 10^{11}$ to $\sim 10^{12} M_\odot$. Are there simple assumptions that will allow extension of the model considered here to more massive galaxies including quenching, and to less massive galaxies including satellites?
\end{itemize}

Even without these modifications, we are finding it useful to compare outputs from the simple model described here to those from a semi-analytic model \citep{Porter+2014,Brennan+2015} run on the same Bolshoi-Planck halos.  We are also comparing the model with zoom-in hydrodynamic galaxy simulations  \citep[such as][]{Zolotov+2015}.  We expect that such comparisons will help to improve both SAMs and simulations as well as this simple model.

\vskip 0.2in

\section*{Acknowledgments} 
We have benefitted from stimulating discussions with Vladimir Avila-Reese,
Andi Burkert, Avishai Dekel, Jerome Fang, John Forbes, Yicheng Guo, Andrew Hearin, Doug Hellinger, Anatoly Klypin, David Koo, Christoph Lee, Nir Mandelker, Rachel Somerville, Frank van den Bosch, and Doug Watson.  We thank Avishai Dekel and Nir Mandelker for detailed comments on an earlier draft of this paper.
ARP was supported by UC-MEXUS Fellowship. PB was supported
by a Giacconi Fellowship from the Space Telescope Science Institute, which is operated 
by the Association of Universities for Research in Astronomy, Incorporated, under NASA contract NAS5-26555.
SF acknowledges support from NSF grant AST-08-08133.
This work was partially supported by grants from UC-MEXUS and HST GO-12060 to the 
CANDELS project, provided by NASA through a grant from the Space Telescope Science Institute, which 
is operated by the Association of Universities for Research in Astronomy,  Incorporated, under NASA 
contract NAS5-26555.  
Authors Behroozi and Primack also benefitted from participation in the 2014 workshop on the galaxy-halo connection at the Aspen Center for Physics organized by Frank van den Bosch and Risa Wechsler.
We thank Yu Lu for providing us with an electronic form of hist data for the mass loading factors. 
We also thank to the anonymous Referee for a constructive report that helped to improve this paper.

\bibliographystyle{mn2efix.bst}
\bibliography{Bibliography}

\end{document}